\documentclass[final,leqno]{siamltex}

\usepackage{mathtools}
\usepackage{amssymb}
\usepackage{wasysym}
\usepackage{graphicx}
\usepackage{stmaryrd}
\usepackage{color}
\usepackage{pdflscape}
\usepackage{epsfig}
\usepackage{graphicx}
\usepackage{epstopdf}
\usepackage{multirow}
\usepackage{hhline}
\usepackage{txfonts}
\usepackage{mathrsfs}
\usepackage{color}
\usepackage{pifont}
\usepackage[dvipsnames]{xcolor}
\usepackage{tikz}
\usepackage[colorlinks=true,
 linkcolor=red,
 citecolor=blue,
 filecolor=magenta,
 urlcolor=cyan]{hyperref}
 
\usepackage[margin=0.9in]{geometry}
\usepackage{subcaption}

\usepackage{mathrsfs}
\newtheorem{thm}{Theorem}[section]

\newtheorem{rem}[thm]{Remark}

\newtheorem{assumb}[thm]{Assumption}

\numberwithin{equation}{section}
\newcommand{\tikzcircle}[2][red,fill=red]{\tikz[baseline=-0.5ex]\draw[#1,radius=#2] (0,0) circle ;}%

\title{A Fractional Subgrid-scale Model for Turbulent Flows: 	
	\protect\\ 	 Theoretical Formulation and \textit{a Priori} Study
	\thanks{This work was supported by the AFOSR Young Investigator Program (YIP) (FA9550-17-1-0150) and by the MURI/ARO (W911NF- 15-1-0562) and by the National Science Foundation Award (DMS-1923201) and the ARO Young Investigator Program Award (W911NF-19-1-0444). Computational resources were provided by the Institute for Cyber-Enabled Research (ICER) at Michigan State University. }}

\author{
	Mehdi Samiee
	\footnote{ D\lowercase{epartment of} M\lowercase{echanical} E\lowercase{ngineering} \& D\lowercase{epartment of} C\lowercase{omputational} M\lowercase{athematics}, S\lowercase{cience}, \lowercase{and}, E\lowercase{ngineering},	
		M\lowercase{ichigan} S\lowercase{tate} U\lowercase{niversity}, 428 S S\lowercase{haw} L\lowercase{ane}, E\lowercase{ast} L\lowercase{ansing}, MI 48824, USA}
	, Ali Akhavan-Safaei
	\footnote{D\lowercase{epartment of} M\lowercase{echanical} E\lowercase{ngineering} \& D\lowercase{epartment of} C\lowercase{omputational} M\lowercase{athematics}, S\lowercase{cience}, \lowercase{and}, E\lowercase{ngineering},	
		M\lowercase{ichigan} S\lowercase{tate} U\lowercase{niversity}, 428 S S\lowercase{haw} L\lowercase{ane}, E\lowercase{ast} L\lowercase{ansing}, MI 48824, USA}
	, and Mohsen Zayernouri
	\footnote{D\lowercase{epartment of} M\lowercase{echanical} E\lowercase{ngineering} \& D\lowercase{epartment of} S\lowercase{tatistics} \lowercase{and} P\lowercase{robability},	M\lowercase{ichigan} S\lowercase{tate} U\lowercase{niversity}, 428 S S\lowercase{haw} L\lowercase{ane}, E\lowercase{ast} L\lowercase{ansing}, MI 48824, USA,  C\lowercase{orresponding author; zayern@msu.edu}}
}

\begin{document}

\maketitle

\begin{abstract}
Coherent structures/motions in turbulence inherently give rise to intermittent signals with sharp peaks, heavy-skirt, and skewed distributions of velocity increments, highlighting the non-Gaussian nature of turbulence. That suggests that the spatial nonlocal interactions cannot be ruled out of the turbulence physics. Furthermore, filtering the Navier-Stokes equations in the large eddy simulation of turbulent flows would further enhance the existing nonlocality, emerging in the corresponding subgrid scale fluid motions.
That urges the development of new nonlocal closure models, which respect the corresponding non-Gaussian statistics of the subgrid stochastic motions. To this end and starting from the filtered Boltzmann equation, we model the corresponding equilibrium distribution function with a \textit{L\'evy}-stable distribution, leading to the proposed fractional-order modeling of subgrid-scale stresses. We approximate the filtered equilibrium distribution function with a power-law term, and derive the corresponding filtered Navier-Stokes equations. Subsequently in our functional modeling, the divergence of subgrid-scale stresses emerges as a single-parameter fractional Laplacian, $(-\Delta)^{\alpha}(\cdot)$, $\alpha \in (0,1]$, of the filtered velocity field. The only model parameter, i.e., the fractional exponent, appears to be strictly depending on the filter-width and the flow Reynolds number. We furthermore explore the main physical and mathematical properties of the proposed model under a set of mild conditions. Finally, the introduced model undergoes \textit{a priori} evaluations based on the direct numerical simulation database of forced and decaying homogeneous isotropic turbulent flows at relatively high and moderate Reynolds numbers, respectively. Such analysis provides a comparative study of predictability and performance of the proposed fractional model.
\end{abstract}

\begin{keywords}
Fractional Laplacian, large eddy simulation, Maxwell \textit{equilibrium} distribution function, Boltzmann transport equation, second-law of thermodynamics, frame invariance, \textit{a priori} analysis, nonlocality
\end{keywords}
\pagestyle{myheadings}
\thispagestyle{plain}

\section{Introduction}
\label{Sec: Intro}
Due to the remarkable advancements in computational capabilities over the last decades, large eddy simulations (LES) have been introduced as a powerful approach in the computation of turbulent structures in \cite{pope2001turbulent, germano1991dynamic}. In modeling subgrid-scale (SGS) structures in LES of turbulent flows, spatially-filtered representation of a turbulent field is required for \textit{a priori} and \textit{a posteriori} analyses. As a key ingredient in the development of SGS models in turbulent flows, the statistical behavior of small scale motions and their cumulative effects on the evolution of the large scales should be incorporated. In comprehensive studies, including numerical and empirical approaches (see e.g., \cite{schmitt2016stochastic,chevillard2006unified, jaberi1996non, cardoso1996dispersion, meneveau1994statistics}), the intermittent statistical behavior of velocity gradients and the development of anomalously intense fluctuations were investigated. These studies confirmed the non-Gaussian statistics of SGS structures and the existence of intermittency in the inertial sub-range of turbulence. By measuring the Lagrangian velocity of tracer particles in a turbulent flow, Mordant et. al. in \cite{mordant2001measurement} explored the intermittent statistics of probability distribution functions (PDF) of the velocity time increments, which is even more highlighted than the corresponding Eulerian spatial increments. Arn{\'e}odo et. al. in \cite{arneodo2008universal} investigated the intermittency and universality properties of velocity temporal fluctuations in highly turbulent flows by quantitatively comparing experimental and numerical data. They described a stochastic phenomenological modelization in the entire range of scales, using a multifractal description, which links Eulerian and Lagrangian statistics. Recently, Buzzicotti et. al. in \cite{buzzicotti2018effect} performed \textit{a priori} analyses of statistical characteristics of resolved-to-subfilter scale (SFS) energy transfer. They quantified the intermittent scaling of the SFS energy transfer as a function of filtering type and described its non-trivial, anomalous deviations from the classical scaling as a function of cutoff scale. In fact, the anomalous behavior of turbulent small scales monotonically deviates from Gaussianity by enlarging the filter width (see e.g., \cite{li2005origin, gille2000velocity, min1996levy, vincent1991satial, chen1989non}). This means that filtering a turbulent field incorporates nonlocal interactions of SGS motions into the resolved scales, which is reflected in heavy-tailed distributions of velocity increments.


Before any discussion on the various strategies of SGS modeling, the reader is referred to \cite{yang2019predictive, nouri2017self, sagaut2006large, pierce2004progress, pope2001turbulent, cerutti2000spectral, jaberi1999filtered, moin1991dynamic} for more history and background on LES of turbulent flows. The SGS modeling strategies are categorized into (I) functional and (II) structural modelings (see e.g., \cite{sagaut2006large}). In the functional strategies, the closure problem can be expressed in the form of a mathematical operator, which is acting on the mean velocity field. Such turbulence models seek only to generate the net kinetic energy transfer from the resolved to small scales (see e.g., \cite{ cui2004new}). However, structural modeling strategies would approximate the SGS stresses in terms of the filtered velocity field, where the SGS structures and statistical properties are recovered from the resolved scale information. Multifractal modelings were introduced in \cite{yang2017multifractal, porte2000scale} as a structural approach to model the underlying intermittent cascading of energy. More specifically, in a study by Burton and Dahem \cite{burton2005multifractal}, a new approach was presented on the multifractal modeling of subgrid-scale stresses in LES of turbulent flows motivated by \textit{a priori} testing. Subsequently, Rasthofer and Gravemier in \cite{rasthofer2013multifractal} proposed a new method of SGS modeling from a multifractal description of the vorticity field. Regarding the non-Gaussain statistics of small scale motions and nonlocal effects in turbulent flows, Hamilington and Dahem in \cite{hamlington2009reynolds} obtained a nonlocal closure modeling from a new derivation of the rapid pressure strain correlation. Recently, Maltba et. al. \cite{maltba2018nonlocal} presented a new semi-local formulation employing a modified large eddy diffusivity (LED) approach, which retains the accuracy of a fully nonlocal approach. It turns out that in formulating SGS models, standard integer-order operators have commonly been used to mathematically represent the anomalous features of small scale motions.

In addition to the considerable progresses in developing nonlocal models using standard methods, fractional calculus appears to be a mathematical tractable tool to describe anomalous phenomena, manifesting in nonlocal interactions, self-similar structures, sharp peaks, and memory effects (see e.g., \cite{wang2019wellposedness, meerschaert2011stochastic}). It seamlessly generalizes the notion of standard integer-order calculus to its fractional-order counterpart, which leads to a broader class of mathematical models. Cushman and Moroni in \cite{cushman2001statistical} developed a theory for modeling anomalous dispersion, which relied on the intermediate scattering function. In another experimental work in \cite{moroni2001statistical}, they obtained the intermediate scattering function using the Lagrangian trajectories for a conservative tracer in a porous medium. Based on the anomalous characteristics of fluctuation processes in turbulence, several studies were conducted to explore the nonlocal modeling of turbulent flows. Chen in \cite{chen2006speculative} proposed a fractional Laplacian stochastic equation to describe intermittent cascade of fully-developed turbulence. Furthermore, Churbanov and Vabishchevich presented a new fractional model to describe turbulent fluid flows in a rectangular duct,  \cite{churbanov2016numerical}. Recently, Egolf and Hutter \cite{egolf2017fractional} proposed nonlocal turbulent models in the form of fractional operators to generalize Reynolds shear stresses in local zero-equation models. Epps and Cushman-Roisin in \cite{epps2018turbulence} derived the Navier-Stokes (NS) equations with fractional Laplacian starting from the Boltzmann transport equation. In their study, they modeled large displacements of fluid particles by \textit{L\`evy} $\alpha$-stable distributions \cite{meerschaert2011stochastic}. Moreover, great progresses have been made towards the theories and numerical solutions to fractional partial differential equations (FPDEs). Samiee et al., \cite{samiee2017Unified, samiee2017unified1} developed a unified Petrov–Galerkin spectral method for a class of FPDEs with two-sided derivatives employing the so-called \textit{Jacobi poly-fractonomials}. Zayernouri and Karniadakis in \cite{zayernouri2013fractional} introduced \textit{Jacobi poly-fractonomials} as a new family of basis/test functions, which are the explicit eigenfunctions of fractional Strum-Liouville problems in bounded domains of the first and second kind. Zhou et. al. developed two efficient first- and second-order implicit-explicit (IMEX) methods for accurate time-integration of stiff/nonlinear fractional differential equations with fractional order $\alpha \in (0,1]$ in \cite{zhou2019fast}. The reader is referred to \cite{fu2019finite, suzuki2018automated, naghibolhosseini2018fractional, kelly2018anomalous, varghaei2019vibration, suzuki2016fractional, naghibolhosseini2015estimation} and the references given therein for more details on fractional modeling of anomalous transport.





In comparison with the recent advances in  SGS modeling of non-Gaussian features, the development of fractional transport modeling of turbulent structures is still at its very early stage. In the present work, we aim to open up a new perspective to functional modeling of the SGS stresses, employing the fractional calculus. This approach implies that we never question the correctness of Navier-Stokes (NS) subject to the Newtonian assumption. Starting from the Boltzmann transport equation, we propose to approximate the filtered Maxwellian equilibrium distribution function of velocity with a \textit{L\'evy} $\alpha$-stable distribution. Accordingly, we derive the filtered NS equations, in which the divergence of SGS stresses is approximated by the fractional Laplacian of filtered velocity field. From a physical point of view, (any generic) filtering the flow field in LES would further contribute to the nonlocal effects, which are rigorously modeled to appear as a fractional Laplacian term in the filtered NS equations. Certainly, by decreasing the filter width, the super-diffusive fractional operator gradually vanishes in compliance with the induced nonlocality.
Here, we briefly highlight the main contributions of this work as follows:


\vspace{0.05 in}
\noindent $\bullet$ We develop a new functional approach to model the SGS stresses by employing the fractional Laplacian of the filtered velocity within the Boltzmann transport framework. The fractional exponent in the model arises from the heavy-tailed behavior of the SGS stresses.

\vspace{0.05 in}
\noindent $\bullet$ We show that the model is frame invariant and constrain it to a set of conditions to preserve the second-law of thermodynamics.

\vspace{0.05 in}
\noindent $\bullet$ We perform the \textit{a priori} studies to assess performance of the model primarily by the correlation and regression coefficients utilizing the results of direct numerical simulation (DNS) for three-dimensional forced and decaying homogeneous isotropic turbulence (HIT) problems. We also investigate the nonlocality of proposed model, as a hallmark of fractional operators, in a range of filter widths.

The paper is organized as follows: in section 2, we introduce some preliminaries on fractional calculus. We outline the problem and discuss the governing equations in section 3. In section 4, we develop the fractional model from the Boltzmann transport equation and study its mathematical and physical properties. In section 5, we provide the details of \textit{a priori} analysis for three-dimensional forced and decaying HIT and study performance of the proposed fractional model. Finally, we summarize the findings with conclusion.

%
%
%
%
%
%

\section{Preliminaries on Fractional Laplacian}
\label{Sec: Notation}
%
%
For modeling SGS stresses in isotropic turbulent flows, the heavy-tailed behavior of \textit{L\'evy} $\alpha$-stable distributions are highly in demand due to their success in capturing singularities and modeling anomalous phenomena (see e.g., \cite{schmitt2016stochastic}).
From the stochastic point of view, the dynamics of the SGS features, which is modeled by an isotropic \textit{L\'evy} $\alpha$-stable process at the microscopic level, can be upscaled by a fractional Laplacian operator. Such operator provides a rigorous tool for the mathematical modeling of nonlocal phenomena \cite{lischke2018fractional}. We denote by $(-\Delta)^{\alpha}$ the fractional Laplacian with $0 < \alpha \leq 1$,
\begin{eqnarray}
\label{FL-1}
(-\Delta)^{\alpha} u(\boldsymbol{x}) &=& \frac{1}{(2 \pi)^d} \int_{\mathbb{R}^d } \vert \boldsymbol{\xi} \vert^{2\alpha} \big {(} u,\, e^{-\mathfrak{i} \boldsymbol{\xi}\cdot \boldsymbol{x} } \big {)}_{L^2} e^{\mathfrak{i} \boldsymbol{\xi}\cdot \boldsymbol{x} } d\boldsymbol{\xi}
\nonumber
\\
&=& \mathcal{F}^{-1} \Big {\{}   \vert \boldsymbol{\xi} \vert^{2\alpha} \mathcal{F}  \big {\{}u\big {\}} (\boldsymbol{\xi}) \Big {\}},
\end{eqnarray}
where $\mathcal{F}$ and $\mathcal{F}^{-1}$ represent the Fourier and inverse Fourier transforms for a real-valued vector $\boldsymbol{\xi}=\xi_j$, $j=1,\, 2,\, 3$, respectively, and $\mathfrak{i}$ denotes the imaginary unit. Moreover, $ ( \cdot,\, \cdot )_{L^2}$ denotes the $L^2$-inner product on $\mathbb{R}^d$, $d=1,2,3$. The Fourier transform of the fractional Laplacian is then obtained as
\begin{equation}
\label{FL-2}
\mathcal{F} \Big {\{} (-\Delta)^{\alpha} u(\boldsymbol{x})  \Big {\}}=\vert \boldsymbol{\xi} \vert^{2\alpha} \mathcal{F}  \big {\{}u\big {\}} (\boldsymbol{\xi}).
\end{equation}
It is worth noting that the integer-order Laplacian is recovered when $\alpha=1$. Considering the definition of $\alpha$-Riesz potential as
\begin{eqnarray}
\label{FL-3-2}
\mathcal{I}_{\alpha} u(\boldsymbol{x}) &=& C_{d,-\alpha} \, \int_{\mathbb{R}^d }\frac{u(\boldsymbol{x})-u(\boldsymbol{s})}{\vert \boldsymbol{x}-\boldsymbol{s}\vert^{d-2\alpha}} ds,
\end{eqnarray}
the fractional Laplacian can also be expressed in the integral form as 
\begin{eqnarray}
\label{FL-3}
(-\Delta)^{\alpha} u(\boldsymbol{x}) &=& C_{d,\alpha} \, \int_{\mathbb{R}^d }\frac{u(\boldsymbol{x})-u(\boldsymbol{s})}{\vert \boldsymbol{x}-\boldsymbol{s}\vert^{2\alpha+d}} d\boldsymbol{s},
\end{eqnarray}
where $C_{d,\alpha} = \frac{2^{2\alpha} \Gamma(\alpha+d/2)}{\pi^{d/2} \Gamma(-\alpha)}$ for $2\alpha \in (0,d)$ and $\Gamma(\cdot)$ represents Gamma function (see \cite{pozrikidis2016fractional}). The $\alpha$-Riesz potential is also formulated in \cite{stein2016singular} as
\begin{eqnarray}
\label{FL-3-3}
\mathcal{I}_{\alpha} u(\boldsymbol{x}) &=& (-\Delta)^{-\alpha} u(\boldsymbol{x}) = \mathcal{F}^{-1} \Big {\{}   \vert \boldsymbol{\xi} \vert^{-2\alpha} \mathcal{F}  \big {\{}u\big {\}} (\boldsymbol{\xi}) \Big {\}}.
\end{eqnarray}
Considering \eqref{FL-3-3}, the Riesz transform is then given by
\begin{eqnarray}
\label{FL-3-4}
\mathcal{R}_j u(\boldsymbol{x}) &=&\nabla_j \, \mathcal{I}_{1} u(\boldsymbol{x}) = \mathcal{F}^{-1} \Big {\{}  -\frac{\mathfrak{i}\xi_j}{ \vert \boldsymbol{\xi} \vert} \mathcal{F}  \big {\{}u\big {\}} (\boldsymbol{\xi}) \Big {\}},
\end{eqnarray}
which is dealt with in formulating the SGS stresses in section \ref{Sec: General FPDE}.


\section{Governing Equations}
In the mathematical description of incompressible turbulent flows, the primitive variables, including the velocity and the pressure fields are represented by $\boldsymbol{V}(\boldsymbol{x},t)=(V_1,\, V_2,\, V_3)$ and $p(\boldsymbol{x},t)$ for $\boldsymbol{x}=x_i$ and $i=1,2,3$, respectively. In the following, the flow field variables are governed by the continuity and the Navier-Stokes (NS) equations, given as
\begin{eqnarray}
\label{GE-1-2}
\frac{\partial V_i}{\partial t}+\frac{\partial V_i\,V_j}{\partial x_j}&=&-\frac{1}{\rho}\frac{\partial p}{\partial x_i}+\frac{1}{\rho}\frac{\partial \sigma_{ij}}{\partial x_j}, \quad i,j=1,2,3,
\end{eqnarray}
where $\rho$ denotes the density and the viscous stress tensor $\sigma_{ij}$ is defined as $ \sigma_{ij}= \mu \, (\frac{\partial V_i}{\partial x_j}+\frac{\partial V_j}{\partial x_i}),$ in which $\mu$ represents the dynamic viscosity for a Newtonian fluid.

In the LES of turbulent flows, the fluid motions are resolved down to some prescribed length scale, filter width ($\mathcal{L}$), which decomposes the velocity field, $\boldsymbol{V}$, into the filtered (resolved), $\boldsymbol{\bar{V}}$, and the residual, $\boldsymbol{v}$, components. The filtered velocity field is obtained by convolution, where $\boldsymbol{\bar{V}}= G \ast \boldsymbol{V}$ and $G = G(\boldsymbol{x})$ denotes the kernel of spatial filtering in the convolution (see \cite{pope2001turbulent, germano1992turbulence}). To produce the filtered velocity field and the true values of SGS stresses, we can adopt any generic isotropic filtering technique.
Accordingly, the filtered NS equations in the index form are derived as
\begin{eqnarray}
\label{GE-2}
\frac{\partial \bar{V}_i}{\partial t}+\frac{\partial \bar{V}_i\,\bar{V}_j}{\partial x_j}&=&-\frac{1}{\rho}\frac{\partial \bar{p}}{\partial x_i}+\frac{\partial }{\partial x_j} (\nu \frac{\partial \bar{V}_i}{\partial x_j}) -\frac{\partial \mathcal{T}^{R}_{ij}}{\partial x_j },
\end{eqnarray}
where the kinematic viscosity is represented by $\nu$ and the SGS stress tensor, $\mathcal{T}^{R}_{ij}=\overline{V_i  V_j}-\bar{V}_i\bar{V}_j$, which must be closed in terms of the filtered flow variables. As the most popular eddy-viscosity closure, we exemplify the Smagorinsky model (SMG), introduced in \cite{smagorinsky1963general}. The SGS stresses in the SMG are modeled by $\mathcal{T}^{R}_{ij} = -2 \nu_{sgs} \bar{S}_{ij}$, where $\bar{S}_{ij}=\frac{\partial \bar{V}_i}{\partial x_j}+\frac{\partial \bar{V}_j}{\partial x_i}$, $\nu_{sgs}= (C_s \mathcal{L})^2\,\vert \bar{\boldsymbol{S}} \vert$, and $\vert \bar{\boldsymbol{S}} \vert = \sqrt{2\bar{S}_{ij}\bar{S}_{ij}}$.


\section{Mathematical Framework}
\label{Sec: General FPDE}

Boltzmann-based frameworks offer a great potential in building transport models at the microscopic level due to their inherent simple mechanism in simulating the interactions of fluid particles through streaming and collision operators. In this section, we develop a new framework to reconcile closure modeling in the Boltzmann transport and the NS equations. Such framework is of great scientific importance specifically for giving a kinetic statistical description of turbulent motions.
\subsection{\textbf{Boltzmann Transport Equation (BTE)}}
\label{sec 4.1}
The kinetic theory aims to describe the motion of particles in a gas from a microscopic point of view. The state of the gas is obtained by a distribution function $f(t,\boldsymbol{x},\boldsymbol{u})$ in the particle phase space such that $f(t,\boldsymbol{x},\boldsymbol{u})\,d\boldsymbol{x}d\boldsymbol{u}$ is defined as the mass of gas particles in phase space within volume $d\boldsymbol{x}d\boldsymbol{u}$ centered on $\boldsymbol{x}$, $\boldsymbol{u}$ at time $t$, where $\boldsymbol{x}$ and $\boldsymbol{u}$ represent gas particle's location and speed, respectively. We note that $\boldsymbol{x}$, $\boldsymbol{u}$, and $t$ are independent variables. Let $f = f(t,\boldsymbol{x},\boldsymbol{u}): \mathbb{R}^{+} \times \mathbb{R}^d \times \mathbb{R}^d \rightarrow \mathbb{R}$ (see e.g., \cite{epps2018turbulence,soto2016kinetic}). Without loss of generality, we take $d=3$. Then ensemble-averaged macroscopic flow variables are given by:
\begin{eqnarray}
\label{GE-4}
\rho &=& \int_{\mathbb{R}^d} f(t,\boldsymbol{x},\boldsymbol{u}) d\boldsymbol{u},
\\
\label{GE-4-2}
V_i&=&\frac{1}{\rho} \int_{\mathbb{R}^d} u_i \, f(t,\boldsymbol{x},\boldsymbol{u}) d\boldsymbol{u}, \quad i=1,2,3,
\end{eqnarray}
where $\rho$ and $V_i$ denote the fluid density and the $i$-th component of flow velocity field in the NS equations, respectively. The accurate description of non-reacting ideal gas particles is governed by the Boltzmann transport equation \cite{soto2016kinetic,succi2001lattice}, which is written as 
\begin{equation}
\label{GE-5}
\frac{\partial f}{\partial t} + \boldsymbol{u}\cdot \nabla f = \big ( \frac{\partial f}{\partial t} \big )_{coll} \equiv \frac{f-f^{eq}}{\tau},
\end{equation}
where $f^{eq}= f^{eq}(t,\boldsymbol{x},\boldsymbol{u})$ represents the equilibrium distribution function and $\tau$ is the relaxation time, which is the required time for fluid particles to reach equilibrium state. The left-hand side  of \eqref{GE-5} represents the streaming of \textit{non-interacting} particles and the right-hand side expresses the collision term due to two-particle interactions \cite{soto2016kinetic}. Assuming that the system of gaseous particles is in thermodynamic equilibrium, the equilibrium distribution is given by the Maxwell distribution \cite{sone2012kinetic}, 
\begin{equation}
\label{GE-6}
f^{eq}(\Delta) = \frac{\rho}{U^3}F(\Delta),
\end{equation}
where $F(\Delta)=e^{-\Delta/2}$, $\Delta=\frac{\vert \boldsymbol{u}-\boldsymbol{V}\vert^2}{U^2}$ and $U$ denotes the agitation speed. For instance, for air at the room-temperature $T=15^{\circ} C$, we get $U=\sqrt{3\textit{k}_B T/m}=502 \, m/s$, in which $\textit{k}_B$, $T$, and $m$ represent the Boltzmann constant, room temperature, and the molecular weight of air \cite{epps2018turbulence}, respectively. It should be pointed out that $F(\Delta)$ is an isotropic function with respect to the velocity variables. Moreover, we define $L$ as the macroscopic characteristic length, $\textit{l}$ as the microscopic characteristic length associated with the Kolmogorov length scale, and $\lambda$ as the mean-free path, which is the average distance, traveled by a particle between successive collisions. Let take $\boldsymbol{x}^{\prime}$ the location of particles before scattering, where $\boldsymbol{x}$ is the current location. Then, $\boldsymbol{x}^{\prime}=\boldsymbol{x}-\delta \boldsymbol{x}$ and $\delta \boldsymbol{x} = (t-t^{\prime})\boldsymbol{u}$, in which we assume that $\boldsymbol{u}$ remains constant during $t-t^{\prime}$. As discussed in \cite{epps2018turbulence}, the analytical solution for the mass probability distribution function is
\begin{equation}
\label{GE-6-2}
f(t,\boldsymbol{x},\boldsymbol{u}) = \int_{0}^{\infty} e^{-s} \,f^{eq}(t-s\tau,\boldsymbol{x}-s\tau \boldsymbol{u},\boldsymbol{u}) \,  ds=\int_{0}^{\infty} e^{-s} f^{eq}_{s,s}(\Delta) ds,
\end{equation}
where $s\equiv\frac{t-t^{\prime}}{\tau}$ and $f^{eq}_{s,s}(\Delta)=f^{eq}(t-s\tau,\boldsymbol{x}-s\tau \boldsymbol{u},\boldsymbol{u})$. 
To establish a mathematical framework for deriving the NS equations from the Boltzman equation in \eqref{GE-5}, we restrain our attention to some necessary assumptions, following \cite{epps2018turbulence}. 
\begin{assumb}
	\label{Rem-1}
	The underlying assumptions for deriving the NS equations from BTE are: 
	
	\noindent $\bullet$ The density, $\rho$, and the thermal agitation, $U$, speed are constant,
	
	\noindent $\bullet$ $s \sim \mathcal{O}(1)$,
	
	\noindent $\bullet$  The mean flow velocity is less than the thermal agitation speed, i.e., $\vert V \vert  \ll U$,
	
	\noindent $\bullet$ $\lambda \ll l \ll L$ and $\tau \ll \frac{L}{\vert \bar{\boldsymbol{V}} \vert}$.
\end{assumb}

\subsection{\textbf{Filtered Boltzmann Transport Equation (FBTE)}}
\label{sec FBTE}

To proceed for the LES of a turbulent flow, we can decompose $f$ to the filtered (resolved) and the residual (unresolved) values as $f = \bar{f} + f^{\prime}$. Recalling from section 3 that overbar represents the spatial isotropic filtering, i.e. $\bar{f}= G \ast f$, where $G$ is the kernel of spatial filtering with the filter width, $\mathcal{L}$. Then, we formulate the filtered BTE (FBTE) according to
\begin{equation}
\label{GE-7}
\frac{\partial \bar{f}}{\partial t} + \boldsymbol{u}\cdot \nabla \bar{f} = \frac{\bar{f}-\overline{f^{eq}(\Delta)}}{\tau}.
\end{equation}
We also define $\bar{\Delta}:=\frac{\vert \boldsymbol{u}-\bar{\boldsymbol{V}} \vert^2}{U^2}$. Following \eqref{GE-6-2}, we obtain the corresponding analytical solution to \eqref{GE-7} in terms of $\overline{f^{eq}(\Delta)}$ as
\begin{equation}
\label{GE-7-2}
\bar{f} (t,\boldsymbol{x}, \boldsymbol{u}) =\int_{0}^{\infty} e^{-s} \,  \overline{f^{eq}_{s,s}(\Delta)}\, ds,
\end{equation}
where $\overline{f^{eq}_{s,s}(\Delta)}=\overline{f^{eq}_{}\big (\Delta(t-s\tau,\boldsymbol{x}-s\tau \boldsymbol{u},\boldsymbol{u}) \big )}$.

It is well-known that the nonlinear term is responsible for the transfer of kinetic energy in the cascade of turbulent kinetic energy from large to small scale turbulent motions. In principle, the SGS stresses originate from the convection term in the NS equations. It accordingly appears natural to recognize the advection term, $ \boldsymbol{u}\cdot \nabla f$, in \eqref{GE-5} as the resource of turbulent motions. Considering \eqref{GE-7-2}, the streaming and collision terms in \eqref{GE-7} can be revised in terms of $\overline{f^{eq}_{s,s}(\Delta)}$, which plays a key role in the development of a model for the SGS stresses. More specifically, the effects of highly vortical flow field on the filtered shifted equilibrium, $\overline{f^{eq}_{s,s}(\Delta)}$, is manifesting in the advection term in \eqref{GE-7}, which gives rise to the SGS stresses. Clearly, the way we treat $\overline{f^{eq}_{s,s}(\Delta)}$ can lead us to the development of a closure model in the LES of turbulent flows. It is very important to note that $\overline{f^{eq}(\Delta)}$ is not equal to the Gaussian distribution of $\bar{\Delta}$, i.e., 
\begin{equation}
\label{GE-8}
\overline{f^{eq}(\Delta)} = \frac{\rho}{U^3}\overline{e^{-\Delta/2}} \neq \frac{\rho}{U^3} e^{-\bar{\Delta}/2}=f^{eq}(\bar{\Delta}).
\end{equation}

A common practice in dealing with $\overline{f^{eq}(\Delta)}$ is to follow the eddy-viscosity approach (see e.g., \cite{girimaji2007boltzmann, premnath2009dynamic, xia2015comparisons}). Generally, the residual scale motions can be modeled by approximating the collision term through a modified relaxation time, $\tau^{\star}$. In an analogy with the standard Smagorinsky SGS model, the Boltzmann equation with the modeled collision term  \cite{girimaji2007boltzmann,sagaut2010toward} can be proposed as 
\begin{equation}
\label{GE-9}
\frac{\partial \bar{f}}{\partial t} + \boldsymbol{u}\cdot \nabla \bar{f} = \frac{\bar{f}-f^{eq}(\bar{\Delta})}{\tau^{\star}},
\end{equation}
where $\tau^{\star}$ accounts for the difference between $\overline{f^{eq}(\Delta)}$ and $f^{eq}(\bar{\Delta})$. Interestingly, $\tau^{\star}$ is inherently associated with the turbulent viscosity in the Smagorinsky model.

Here, we outline a new framework to develop an LES modeling strategy from the BTE using a non-Gaussian stochastic process. Without loss of generality, we consider $G(\boldsymbol{r}) = \frac{1}{\mathcal{L}} H(\frac{1}{2}\mathcal{L}-\vert \boldsymbol{r} \vert )$ as the convolution kernel of box filtering, where $H(\cdot)$ denotes a Heaviside step function. Therefore, 
\begin{equation}
\label{GE-9-3}
\overline{f^{eq}(\Delta)} =  G \ast f^{eq}\big (\Delta(t,\boldsymbol{u},\boldsymbol{x})\big ) =  \int_{R_f}^{} G(\boldsymbol{r}) \,  f^{eq}\big (\Delta(t,\boldsymbol{u},\boldsymbol{x}-\boldsymbol{r})\big ) \, d\boldsymbol{r},
\end{equation}
where $R_f=[-\frac{\mathcal{L}}{2},-\frac{\mathcal{L}}{2}]^3$. Technically, $\overline{f^{eq}(\Delta)} $ represents a summation of exponential functions, which leads to its multi-exponential characteristics especially when $\mathcal{L}$ gets increased. That is, by enlarging $\mathcal{L}$, we are incorporating more information into $\overline{f^{eq}(\Delta)} $ according to \eqref{GE-9-3}, which essentially induces more heaviness to the statistical behavior of $\overline{f^{eq}(\Delta)}$. Thus, $\overline{f^{eq}(\Delta)}$ deviates more and more from the Gaussianity of $f^{eq}(\Delta)$  (see e.g., \cite{buzzicotti2018effect,  shtrauss2007decomposition, li2005origin}). It should be noted that we are permitted to employ any generic type of filtering.

For the purpose of modeling $\overline{f^{eq}(\Delta)}$, it is understood from \cite{rubinstein2003polymer, chu2010power} that the multi-exponential distributions can be fitted with a power-law model, in which the discrepancy between the model and true values can be reduced by increasing the number of exponential functions. Accordingly, we propose to model $\overline{f^{eq}(\Delta)}-f^{eq}(\bar{\Delta})$ with a power-law distribution, which follows as
\begin{equation}
\label{GE-9-2}
\overline{f^{eq}(\Delta)} - f^{eq}(\bar{\Delta}) \simeq f^{Model}(\bar{\Delta}) =  \mathfrak{C}_{\beta}\, f^{\beta}(\bar{\Delta}),
\end{equation}
where $f^{\beta}(\bar{\Delta}) = \frac{\rho}{U^3}F^{\beta}(\Delta)$, in which $F^{\beta}(\Delta)$ denotes an isotropic \textit{L\'evy} $\beta$-stable distribution. We assume $ \mathfrak{C}_{\beta}$ is a real-valued constant number. Moreover, we consider $\beta\in (\frac{d}{2},1+\frac{d}{2})$ and $d=3$ represents the dimension of physical domain.

\begin{rem}
	\label{Rem-2}
	Unlike the fractional exponent in \cite{epps2018turbulence}, $\beta$ relies not only on the thermodynamic properties and boundary conditions, but also it is a function of Taylor Reynolds number, $Re_{\lambda}$ (defined further in Table \ref{Table: properties}), and $\mathcal{L}$. It is also worth mentioning that the power-law distribution can be well-suited in the modeling of multi-exponential functions if the filter width is chosen large enough to incorporate nonlocal interactions. 
\end{rem}

Therefore, we propose to model $\overline{f^{eq}(\Delta)}$ in the collision term by using an isotropic \textit{L\'evy} $\beta$-stable distribution. Therefore, the FBTE is approximated by
\begin{equation}
\label{GE-12}
\frac{\partial \bar{f}}{\partial t} + \boldsymbol{u}\cdot \nabla \bar{f} = \frac{\bar{f}-f^{eq}(\bar{\Delta})+f^{eq}(\bar{\Delta})-\overline{f^{eq}(\Delta)}}{\tau} \simeq \frac{\bar{f}-f^{eq}(\bar{\Delta})-f^{Model}(\bar{\Delta})}{\tau}.
\end{equation}
For the sake of simplicity, we take $f^{*}(\bar{\Delta}) = f^{eq}(\bar{\Delta})+f^{Model}(\bar{\Delta})$. In comparison to the eddy-viscosity models, we approximate the collision term by replacing $\overline{f^{eq}(\Delta)}$ by $f^{*}(\bar{\Delta})$ rather than modifying $\tau$, which leverages incorporating nonlocal interactions in turbulent flows.

\subsection{\textbf{Derivation of the FSGS model }}
The macroscopic continuum variables, associated with \eqref{GE-2}, can be expressed in terms of filtered distribution function in \eqref{GE-12} as
\begin{eqnarray}
\label{GE-13}
\bar{\rho} &=& \int_{\mathbb{R}^d} \bar{f}(t,\boldsymbol{x},\boldsymbol{u}) d\boldsymbol{u},
\\
\bar{V}_i&=&\frac{1}{\rho} \int_{\mathbb{R}^d} u_i \, \bar{f}(t,\boldsymbol{x},\boldsymbol{u}) d\boldsymbol{u}, \quad i=1,2,3,
\end{eqnarray}
where $\rho = \bar{\rho}$ for an incompressible flow. It follows from \cite{epps2018turbulence, pottier2010nonequilibrium} that by multiplying both sides of \eqref{GE-12} by a collisional invariant $\mathcal{X}=\mathcal{X}(\boldsymbol{u})$ and then integrating over the kinetic momentum, we attain
\begin{equation}
\label{GE-13-2}
\int_{\mathbb{R}^d} \mathcal{X} \big ( \frac{\partial \bar{f}}{\partial t} + \boldsymbol{u}\cdot \nabla \bar{f}\big ) d\boldsymbol{u} = \int_{\mathbb{R}^d} \mathcal{X} \big ( \frac{\bar{f}-f^{*}(\bar{\Delta})}{\tau} \big ) d\boldsymbol{u},
\end{equation}
where the choices of $\mathcal{X} = 1, \, \boldsymbol{u}$ lead to the conservation of mass and momentum equations, respectively. As noted in \cite{saint2009hydrodynamic}, due to the microscopic reversibility of the particles (the collisions are taken to be elastic), $\int_{\mathbb{R}^d} \mathcal{X} \big ( \frac{\bar{f}-f^{*}(\bar{\Delta})}{\tau} \big ) d\boldsymbol{u} = 0$. 
This allows \eqref{GE-13-2} to be found as
\begin{eqnarray}
\label{GE-14}
\int_{\mathbb{R}^d}  \Big ( \frac{\partial \bar{f}}{\partial t} +  \nabla \cdot (\boldsymbol{u} \bar{f}) \Big ) d\boldsymbol{u} = 0 \quad &\Longrightarrow& \quad \frac{\partial \rho}{\partial t} + \nabla\cdot (\rho \bar{\boldsymbol{V}}) = 0, 
\\
\label{GE-15}
\int_{\mathbb{R}^d} \Big ( \boldsymbol{u}  \frac{\partial \bar{f}}{\partial t} + \nabla \cdot (\boldsymbol{u}^2 \bar{f}) \Big ) d\boldsymbol{u} = 0 \quad &\Longrightarrow& \quad \rho \frac{\partial \bar{\boldsymbol{V}}}{\partial t} + \nabla \cdot \int_{\mathbb{R}^d} \boldsymbol{u}^2 \bar{f} d\boldsymbol{u} = 0,
\end{eqnarray}
in which $\boldsymbol{u}$ is independent of $t$ and $\boldsymbol{x}$. Reminding that the filter convolution kernel, $G=G(\boldsymbol{x})$, is independent of $t$ and $\boldsymbol{u}$ and thereby, Assumption \ref{Rem-1} still holds. In \eqref{GE-15}, by adding and subtracting $\bar{\boldsymbol{V}}\bar{\boldsymbol{V}}$, the advection term, $\boldsymbol{u}^2$, is evaluated as
\begin{eqnarray}
\label{GE-16}
\int_{\mathbb{R}^d} \boldsymbol{u}^2 \bar{f} d\boldsymbol{u} &=& \int_{\mathbb{R}^d} (\boldsymbol{u}-\bar{\boldsymbol{V}})(\boldsymbol{u}-\bar{\boldsymbol{V}}) \bar{f} d\boldsymbol{u}+ \int_{\mathbb{R}^d} \bar{\boldsymbol{V}}\bar{\boldsymbol{V}} \bar{f} d\boldsymbol{u}
\nonumber
\\
&=& \int_{\mathbb{R}^d} (\boldsymbol{u}-\bar{\boldsymbol{V}})(\boldsymbol{u}-\bar{\boldsymbol{V}}) \bar{f} d\boldsymbol{u}+ \rho \bar{\boldsymbol{V}}^2.
\end{eqnarray}
Plugging \eqref{GE-16} into \eqref{GE-15}, we obtain
\begin{equation}
\label{GE-17}
\rho \Big (\frac{\partial \bar{\boldsymbol{V}}}{\partial t} + \nabla \cdot \bar{\boldsymbol{V}}^2 \Big ) = -\nabla \cdot \boldsymbol{\varsigma},
\end{equation}
where 
\begin{equation}
\label{GE_17_2}
\varsigma_{ij}=\int_{\mathbb{R}^d} (u_i-\bar{V}_i)(u_j-\bar{V}_j) \bar{f} d\boldsymbol{u}.
\end{equation} 
It is worth mentioning that the Cauchy and filtered SGS stresses arise from $\varsigma_{ij}$. 
Considering \eqref{GE-7-2}, we formulate $\varsigma_{ij}$ in  \eqref{GE_17_2} as 
\begin{eqnarray}
\label{GE-19-1}
\varsigma_{ij} &=& \int_{\mathbb{R}^d}\int_{0}^{\infty} e^{-s} (u_i-\bar{V}_i)(u_j-\bar{V}_j) f^{*}
_{s,s}(\bar{\Delta}) ds \, d\boldsymbol{u},
\end{eqnarray}
where  $f_{s,s}^{Model} := f^{Model}\big (\bar{\Delta}(t-s\tau,\boldsymbol{x}-s\tau \boldsymbol{u}, \boldsymbol{u}) \big )$, $f_{s,s}^{eq} := f^{eq}\big (\bar{\Delta}(t-s\tau,\boldsymbol{x}-s\tau \boldsymbol{u}, \boldsymbol{u}) \big )$, thus, $f_{s,s}^{*} := f^{*}\big (\bar{\Delta}(t-s\tau,\boldsymbol{x}-s\tau \boldsymbol{u}, \boldsymbol{u}) \big )$. In Appendix, we prove that the temporal shift can be dropped from \eqref{GE-19-1} following the derivations of fractional NS equations in \cite{epps2018turbulence}. Consequently, $\varsigma_{ij}$ in \eqref{GE-19-1} can be simplified to
\begin{eqnarray}
\label{GE-20}
\varsigma_{ij} &=& \int_{\mathbb{R}^d}\int_{0}^{\infty} e^{-s} (u_i-\bar{V}_i)(u_j-\bar{V}_j) f^{eq}
_{s}(\bar{\Delta}) ds \, d\boldsymbol{u} 
\\
\nonumber
&& + \int_{\mathbb{R}^d}\int_{0}^{\infty} e^{-s} (u_i-\bar{V}_i)(u_j-\bar{V}_j) f^{Model}
_{s}(\bar{\Delta}) ds \, d\boldsymbol{u}.
\end{eqnarray}
According to the kinetic definition of static pressure, $p=\rho \, U^2$, we decouple $\varsigma_{ij}$ as
\begin{equation}
\label{GE-21}
\varsigma_{ij}=-\bar{p}\delta_{ij}+\mathcal{T}_{ij},
\end{equation}
where 
\begin{equation}
\label{GE-22}
-\bar{p}\delta_{ij} = \int_{\mathbb{R}^d} (u_i-\bar{V}_i)(u_j-\bar{V}_j) f^{*}(\bar{\Delta}) d\boldsymbol{u} \int_{0}^{\infty} e^{-s} ds
\end{equation}
and $\mathcal{T}_{ij}=\mathcal{T}_{ij}^{Shear}+\mathcal{T}_{ij}^{R}$ denotes the sum of shear stress tensor, $\mathcal{T}_{ij}^{Shear}$, and the SGS stress tensor, $\mathcal{T}_{ij}^{R}$. It is worth noting that in \eqref{GE-22} when $i \neq j$, $(u_i-\bar{V}_i)(u_j-\bar{V}_j) f^{*}(\bar{\Delta})$ represents an odd function of $u_i$ and $u_j$; consequently, $\int_{\mathbb{R}^d} (u_i-\bar{V}_i)(u_j-\bar{V}_j) f^{*}(\bar{\Delta}) d\boldsymbol{u} = 0$. Considering $f^{*}_{s}(\bar{\Delta})=f^{*}_{}(\bar{\Delta})+(f^{*}_{s}(\bar{\Delta})-f^{*}_{}(\bar{\Delta}))$, $\mathcal{T}_{ij}$ is then obtained as
\begin{equation}
\label{GE-24}
\mathcal{T}_{ij}=\int_{0}^{\infty}  \int_{\mathbb{R}^d} (u_i-\bar{V}_i)(u_j-\bar{V}_j) (f^{*}_{s}(\bar{\Delta})-f^{*}(\bar{\Delta})) e^{-s} d\boldsymbol{u}  \, ds. 
\end{equation}
By ascribing the Gaussian distribution $f^{eq}(\bar{\Delta})$ to $\mathcal{T}_{ij}^{Shear}$ and the isotropic \textit{L\'evy} $\beta$-stable distribution, $f^{Model}(\bar{\Delta})$, to $\mathcal{T}_{ij}^{R}$, $\mathcal{T}_{ij}$ in \eqref{GE-24} is decomposed to
\begin{eqnarray}
\label{GE-25}
\mathcal{T}_{ij}^{Shear} &=& \int_{0}^{\infty}  \int_{\mathbb{R}^d} (u_i-\bar{V}_i)(u_j-\bar{V}_j) (f^{eq}_{s}(\bar{\Delta})-f^{eq}(\bar{\Delta})) e^{-s} d\boldsymbol{u}\,  ds,
\\
\label{GE-26}
\mathcal{T}_{ij}^{R} &=& \int_{0}^{\infty}  \int_{\mathbb{R}^d} (u_i-\bar{V}_i)(u_j-\bar{V}_j) (f^{Model}_{s}(\bar{\Delta})-f^{Model}(\bar{\Delta})) e^{-s} d\boldsymbol{u}\,  ds
\nonumber
\\
&=& \int_{0}^{\infty}  \int_{\mathbb{R}^d} (u_i-\bar{V}_i)(u_j-\bar{V}_j) (f^{\beta}_{s}(\bar{\Delta})-f^{\beta}(\bar{\Delta})) e^{-s} d\boldsymbol{u}\,  ds.
\end{eqnarray}

In Appendix, we discuss the evaluations of $\mathcal{T}_{ij}^{Shear}$ and $\mathcal{T}_{ij}^{R}$ in terms of the macroscopic quantities, including $\rho$ and $\bar{\boldsymbol{V}}$. Eventually, the shear stresses are given by
\begin{equation}
\label{GE-26-2}
\mathcal{T}_{ij}^{Shear} = \mu \Big ( \frac{\partial \bar{V}_i}{\partial x_j} + \frac{\partial \bar{V}_j}{\partial x_i}  \Big ),
\end{equation} 
where $\mu=\rho U^2\tau$ denotes the kinematic viscosity. Furthermore, we formulate the divergence of SGS stress tensor as
\begin{equation}
\label{GE-27}
(\nabla \cdot \mathcal{T}^{R})_i=\frac{\rho (U\tau)^{2\alpha}}{\tau}\Gamma(2\alpha+1) \, C_{\alpha} \int_{\mathbb{R}^d} \frac{\bar{V}_i(\boldsymbol{x}^{\prime})-\bar{V}_i(\boldsymbol{x})}{\vert \boldsymbol{x}^{\prime}-\boldsymbol{x} \vert^{2\alpha+d}}d\boldsymbol{x}^{\prime},
\end{equation}
where $\alpha=-\beta-d/2$. Regarding the definition of fractional Laplacian given in \eqref{FL-3}, we can rewrite equation \eqref{GE-27} as
\begin{equation}
\label{GE-28_2}
(\nabla \cdot \mathcal{T}^{R})_i=\mu_{\alpha} (-\Delta)^{\alpha} \bar{V}_i,
\end{equation}
in which $\mu_{\alpha}=\frac{\rho (U\tau)^{2\alpha}}{\tau}\Gamma(2\alpha+1) \, C_{\alpha}$ and $C_{\alpha}=\frac{2^{2\alpha} \Gamma(\alpha+d/2)}{\pi^{d/2} \Gamma(-\alpha)} \mathfrak{C}_{\alpha}$ and $\mathfrak{C}_{\alpha}$ is a real-valued constant. Therefore, the filtered NS equations, developed from the filtered kinetic transport equation, is described by
\begin{eqnarray}
\label{GE-28}
\frac{\partial \bar{V}_i}{\partial t}+\frac{\partial \bar{V}_i\,\bar{V}_j}{\partial x_j}&=&-\frac{1}{\rho}\frac{\partial \bar{p}}{\partial x_i}+\nu  \Delta \bar{V}_i -\nu_{\alpha} (-\Delta)^{\alpha} \bar{V}_i,
\end{eqnarray}
where $\alpha \in (0,1]$, $\nu_{\alpha} = \frac{\mu_{\alpha}}{\rho}$. With a proper choice of $\alpha=\alpha(Re_{\lambda},\mathcal{L})$, in which $\alpha \vert_{\mathcal{L} =0} = 1$, the FSGS model is able to capture the heavy-tailed distribution of the SGS quantities and predict the corresponding high-order statistical moments. By setting $\mathcal{L} = 0$, we obtain $\nu_{\alpha=1} = 0$, and hence $\nu_{\alpha} (-\Delta)^{\alpha} \bar{V}_i = 0$, which evidently recovers the exact NS equations, given in \eqref{GE-1-2}.


\begin{rem}
	\label{Rem-3}
	In \cite{epps2018turbulence}, Epps and Cushman-Roisin evaluated the fractional NS equations from the BTE by replacing $f^{eq}(\Delta)$ as a Gaussian distribution with a \textit{L\'evy} $\beta$-stable distribution and splitting the jumps of particles into small and large scales. From this perspective, the fractional exponent, $\alpha$, in the fractional NS equations is introduced only as a function of fluid properties and boundary conditions. Unlike that, we developed the proposed fractional SGS model from the FBTE by approximating $\overline{f^{eq}(\Delta)}-f^{eq}(\bar{\Delta})$ with a \textit{L\'evy}-$\beta$ stable distribution, in which $\overline{f^{eq}(\Delta)} = G \ast  f^{eq}(\Delta)$ and $G = G(\boldsymbol{x})$. Besides, we found that the factional exponent depends on the flow properties, $Re_{\lambda}$, and also $\mathcal{L}$. Therefore, by setting $\mathcal{L} = 0$ we recover the standard NS equations at any $Re_{\lambda}$. 
\end{rem}

From the Fourier definition of fractional Laplacian and the Riesz transform in Section \ref{Sec: Notation}, it is straightforward to verify that 
\begin{eqnarray}
\mathcal{F} \Big {\{} (-\Delta)^{\alpha} \bar{V}_j \Big {\}} =  \mathfrak{i} \xi_i \Big (  -\frac{\mathfrak{i} \xi_i}{ \vert \boldsymbol{\xi} \vert} \Big ) \, (\vert \boldsymbol{\xi} \vert^2 )^{\alpha-\frac{1}{2}} \mathcal{F} \Big {\{} \bar{V}_j \Big {\}},
\end{eqnarray}
which leads to
\begin{equation}
\label{stss-1}
(-\Delta)^{\alpha} \bar{\boldsymbol{V}} = \nabla_j (\mathcal{R}_j (-\Delta)^{\alpha-\frac{1}{2}} \bar{\boldsymbol{V}}).
\end{equation}
Therefore, we obtain
\begin{equation}
\label{stss-1-2}
\nabla \cdot \mathcal{T}^{R} =  \nabla \cdot (\mathcal{R} (-\Delta)^{\alpha-\frac{1}{2}} \bar{\boldsymbol{V}}).
\end{equation} 
Using \eqref{stss-1-2}, we can find the equivalent form of the SGS stress tensor as
\begin{equation}
\label{stss-2}
\mathcal{T}^{*}_{ij} = \mathcal{T}^{R}_{ij} + C =  \frac{1}{2} (\mathcal{R}_j (-\Delta)^{\alpha-\frac{1}{2}} \bar{V}_i+\mathcal{R}_i (-\Delta)^{\alpha-\frac{1}{2}} \bar{V}_j),
\end{equation}
where $C$ is a real-valued constant. $\mathcal{T}^{*}_{ij}$ is dealt with later in section \ref{Sec: a priori analysis} in the computation of the correlation coefficients .


\begin{rem}
	As described earlier in \eqref{FL-2}, $\mathcal{F} \Big {\{}  (-\Delta)^{\alpha} \bar{\boldsymbol{V}}  \Big {\}}= \vert \boldsymbol{\xi} \vert^{2\alpha} \mathcal{F} \big {\{}\bar{\boldsymbol{V}} \big {\}} $. Similar to the eddy-viscosity models, $\nabla \cdot \mathcal{T}^{R}$ can be explicitly derived in the Fourier domain, hence maintaining the high-order accuracy of scheme. 
\end{rem}

\subsection{\textbf{Physical Properties}}
\label{Solution and Test Function Spaces}
In order to ensure that the developed FSGS model is physically and mathematically consistent with the filtered NS equations, we introduce a mild condition for the model in accordance with the second law of thermodynamics and also examine the frame invariant modeling as follows.

\subsubsection{\textit{Second-law of Thermodynamics}} 
The contribution of filtered momentum equation in the entropy production rate is formulated in \cite{sun2017eddy} as
\begin{equation}
\label{VE-1}
\dot{S}_{prod}=\frac{1}{T} \big ( \mathcal{T}^{Shear}  :  \nabla \bar{\boldsymbol{V}} +  \mathcal{T}^{R}  :  \nabla \bar{\boldsymbol{V}}    \big ),
\end{equation}
where $T$ represents the temperature of flow and ``$:$'' denotes a double dot product operator. In thermodynamic analysis of the exact NS equations, discussed in \cite{bejan2013entropy}, it is proven that $\mu>0$, in the description of $\mathcal{T}_{ij}^{Shear} = \mu \Big ( \frac{\partial \bar{V}_i}{\partial x_j} + \frac{\partial \bar{V}_j}{\partial x_i}  \Big )$. Regarding $\mu>0$ and $\boldsymbol{\mathcal{T}}^{R}=\mathcal{R} (-\Delta)^{\alpha - \frac{1}{2}} \bar{\boldsymbol{V}}$ in \eqref{stss-1-2}, the underlying coefficient in the FSGS model, $\mu_{\alpha}$, should satisfy
\begin{equation}
\label{VE-2}
\mu_{\alpha} \leq \mu \min \left\vert \frac{\nabla \bar{\boldsymbol{V}} : \nabla \bar{\boldsymbol{V}}}{(\mathcal{R} (-\Delta)^{\alpha-\frac{1}{2}} \bar{\boldsymbol{V}}) : \nabla \bar{\boldsymbol{V}}} \right\vert,
\end{equation}
to ensure the positivity of entropy generation rate.

\subsubsection{\textit{Frame Invariance}} 

The SGS stresses and their divergence are separately proven to be frame invariant \cite{oberlack1997invariant, thomas1988frame}, which contribute to invariant characteristics of the NS equations. In order to reproduce all local and nonlocal turbulent solutions in the LES of turbulence, SGS models should undergo certain restrictions to follow such invariant properties; otherwise, the value of turbulent stresses may change with any frame movement. 

It is apparent that in the FSGS model, $\mu_{\alpha}$ is frame invariant. Additionally, as a generator of \textit{L\'evy}-stable processes, the fractional Laplacian operator is proven to be rotationally and Galilean invariant (see \cite{napoli2018symmetry, boling2015fractional}); therefore, we do not need to impose any additional constraint on the FSGS model.





\section{\textit{A Priori} Analysis of the Fractional SGS Model}
\label{Sec: a priori analysis}
We perform \textit{a priori} tests using DNS database to study the performance and capability of the proposed model in capturing anomalous behavior of SGS quantities. To pursue the \textit{a priori} evaluations, we introduce two primary cases: three-dimensional forced and decaying homogeneous isotropic turbulent flows with periodic boundary conditions as follows. 

\subsubsection*{\textbf{Case (I): Forced HIT}}
\label{forced} 
Forced HIT is a canonical benchmark in studying the performance of subgrid-scale models. This test case has the obvious advantage of allowing the statistical features to be approximately stationary. Here, the corresponding computational domain is specified as $\Omega=[0,2\pi]^3$, which is uniformly discretized on a Cartesian grid using $1024^3$ grid points. The Johns Hopkins Turbulence Databases (JHTDB)\footnote[1]{http://turbulence.pha.jhu.edu} has provided public access to DNS database of a forced isotropic homogeneous turbulent flow, which is characterized by the micro-scale statistical properties presented in Table \ref{Table: properties}. For more information, the reader is referred to \cite{JHTDB-1, JHTDB-2}.
\begin{table}[tp]
	\centering
	\caption{Computational parameters and statistical features of a forced HIT problem, provided by JHTDB \cite{JHTDB-1}
	}
	\vspace{-0.1 in}
	\label{Table: properties} 
	\begin{tabular}{c c c c c}
		\hline \hline  
		\vspace{-0.1 in}
		\\
		$Re_{\lambda}=  \frac{u'_{rms} \lambda}{\nu}$
		& $u'_{rms} = \sqrt{\frac{2}{3}E_{tot}} $   & $E_{tot} = \left\langle v'_i \, v'_i \right\rangle $   & $\nu$  & $\varepsilon = 2\nu \left\langle \bar{S}_{ij}\bar{S}_{ij}  \right\rangle $  
		\vspace{0.1 in}
		\\
		&$(m/{sec})$&$(m^2/{sec}^2)$&$ (m^2/{sec})$&$(m^2/{sec}^3) $
		\\
		\cline{1-5}    
		\vspace{-0.1 in}
		\\
		$437$ &  $0.686 $&  $0.93 $ &  $1.85 \times 10^{-4}  $ & $9.28 \times 10^{-2}  $
	\end{tabular}
	\vspace{0.1 in}
\end{table}

In the \textit{a priori} assessments of the FSGS model, the filtered velocity fields are obtained from the DNS data by using a three-dimensional box filtering, in which we set $\mathcal{L}_{\delta}=\frac{\mathcal{L}}{2\delta}=2^j$ for $j=0,\cdots,5$, where $\mathcal{L}$ and $\delta$ represent filter and grid widths, respectively.

%

\subsubsection*{\textbf{Case (II): Decaying HIT}}
\label{decaying}
In terms of \textit{a priori} tests, the DNS of decaying HIT set the ground to evaluate the modeling capabilities of FSGS while $Re_{\lambda}$ experiences a decaying process. Furthermore, the DNS dataset of decaying HIT gives us the opportunity to conduct a series of \textit{a priori} tests to evaluate the performance of the proposed model for a wider range of Reynolds numbers.

Similar to \textbf{Case (I)}, the computational domain is chosen to be the cube of $\Omega=[0,2\pi]^3$ with the periodic boundary conditions. We start from a three-dimensional fully-developed HIT as the initial condition, which was previously obtained from the DNS of a forced HIT. The skewness and flatness of the velocity derivatives for the initial condition data are approximately $-0.5$ and $4.0$, respectively. Table \ref{Table: C2-properties} shows the micro-scale statistical properties of the initial condition, which are described in Table \ref{Table: properties}. It should be mentioned that $\mathscr{L}$ and $\tau_{\mathscr{L}}$ represent the integral length scale and the eddy turnover time, respectively. 
\begin{table}[tp]
	\centering
	\caption{Micro-scale statistical characteristics of turbulence for the applied initial condition in DNS of decaying HIT.}
	\vspace{-0.1 in}
	\label{Table: C2-properties} 
	\begin{tabular}{c c c c c c c c}
		\hline \hline  
		\vspace{-0.1 in}
		\\
		$Re_{\lambda}$ & $u'_{rms}$   & $K$   & $\nu$  & $\varepsilon$ & $\mathscr{L}$ & $\tau_{\mathscr{L}}$
		\vspace{0.1 in}
		\\
		& $(m/{sec})$ & $(m^2/{sec}^2)$ & $ (m^2/{sec})$ & $(m^2/{sec}^3)$ & $(m)$ & $(sec)$
		\\
		\cline{1-7}     
		\vspace{-0.1 in}
		\\
		$66$ &  $0.186 $&  $0.052 $ & 0.001 & $4.17 \times 10^{-3}$ & 0.275 & 1.478
	\end{tabular}
	\vspace{0.1 in}
\end{table}

\begin{figure}[h!]
	\centering
	\begin{subfigure}[b]{0.48\textwidth}
		\centering
		\includegraphics[width=3.0in]{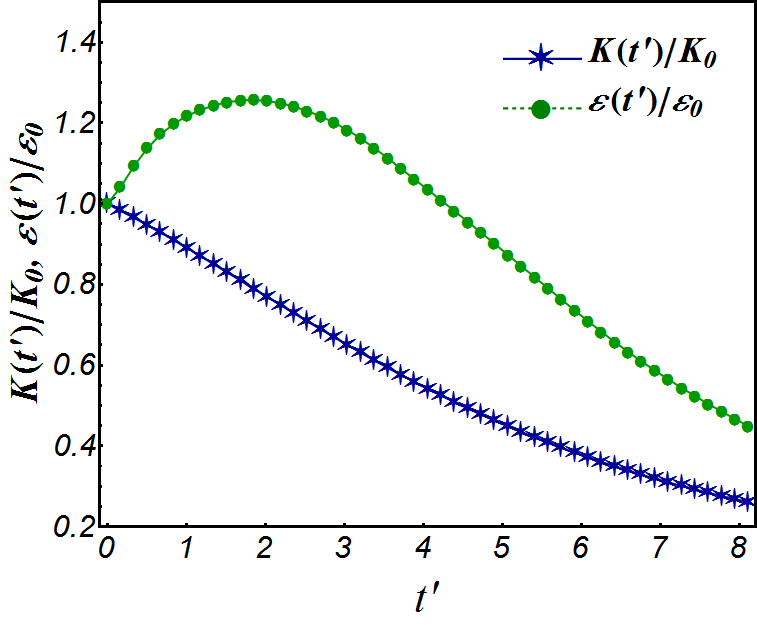}
		\subcaption{\scriptsize Normalized kinetic energy and dissipation rate}
	\end{subfigure}
	\begin{subfigure}[b]{0.48\textwidth}
		\centering
		\includegraphics[width=3.0in]{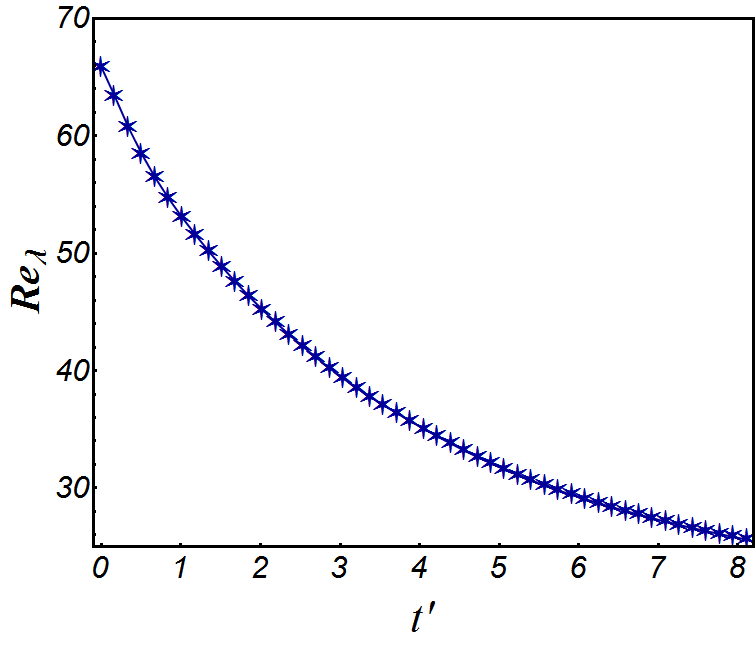}
		\subcaption{\scriptsize Taylor micro-scale Reynolds number}
	\end{subfigure}
	\caption{\label{fig: DNS_decaying}Time evolution of $K(t')/K_0$, $\varepsilon(t')/\varepsilon_0$, and $Re_\lambda(t')$ during the DNS of decaying HIT}
\end{figure}

Further, we conduct the numerical simulation of decaying HIT using the incompressible Navier-Stokes solver of \texttt{NEKTAR++}, which is an open-source spectral/$hp$ element framework \cite{Nektar++2015,Nektar++2019}. Using a $C^0$-continuous Galerkin projection, the discretized domain consists of $64^3$ uniform tetrahedral elements and the fifth-order modified polynomials, $p=5$, as the basis functions within each element. In other words, our computational domain would be a uniformly discretized cube with $256^3$ grid points. The applied solver works based on the velocity-correction method and for time integration we use the second-order IMEX scheme.
Let $k_{max}$ and $\eta=(\nu^3/\varepsilon)^{1/4}$ denote the maximum wave number of turbulence and the Kolmogorov length scale, respectively. As a measure of accuracy, we evaluate $k_{max}\eta>2.6$, which ensures that Kolmogorov scale motions are well-resolved. Figure \ref{fig: DNS_decaying} depicts the time evolution of normalized turbulent kinetic energy, $K(t')/K_0$, normalized dissipation rate, $\varepsilon(t')/\varepsilon_0$, and $Re_\lambda(t')$, where $t'=t/\tau_{\mathscr{L}}$ is the dimensionless time and $K_0=K(t'=0),\ \varepsilon_0=\varepsilon(t'=0),$ and $\tau_{\mathscr{L}}$ are the values reported in Table \ref{Table: C2-properties}.
The kinetic energy is monotonically decaying in Figure \ref{fig: DNS_decaying} while the dissipation rate first experiences an increase up to approximately two large eddy turnover times, $t'\approx 2$, and later monotonically decays. The decay of dissipation rate occurs when the energy spectrum starts to completely decay at the entire wavenumbers, which is consistent with the physics of decaying HIT problems \cite{pope2001turbulent}. To conduct the \textit{a priori} analysis of \textbf{Case (II)}, we collect the velocity field data, starting from $t'\approx 2$ , where $Re_\lambda\approx 45$, and we set $\mathcal{L}_{\delta}=j$ for $j=1,\cdots,15$.

\subsection{\textbf{Estimation of Fractional Exponent $\alpha$}}
To achieve a high degree of accuracy and performance in the FSGS model, the model parameters are considered to be a function of $\mathcal{L}_{\delta}$ and $Re_{\lambda}$. By assuming $\mathfrak{C}_{\alpha}$ as a real-valued function of $\alpha$ in \eqref{GE-9-2}, there is only one adjustable model parameter, $\alpha$, given in \eqref{GE-28}. Conventionally, the correlation and regression coefficients are known as the primary tools in \textit{a priori} tests for tuning the parameters associated with an SGS model. Following \cite{lu2007priori}, we denote by $\varrho_i \in [-1, \, 1]$
and  $\mathcal{R}_i$ for $i=1,2,3$
the correlation and regression coefficients between $ [\nabla \cdot \mathcal{T}^{R}]_i^{DNS}$ from the filtered DNS data and $[\nabla \cdot \mathcal{T}^{R}]_i^{FSGS}$ from the FSGS model, respectively. 
Moreover, the correlation coefficient associated with a component of SGS stresses, $\mathcal{T}^{R}_{ij}$, is indicated by $\varrho_{ij}$ with dual subscripts, where $i,j=1,2,3$. 
Since the FSGS model is strictly limited to access the straight form of SGS stresses, we employ the equivalent SGS stresses, $\mathcal{T}^{*}_{ij}$, given in \eqref{stss-2} to attain $\varrho_{ij}$. Therefore, $\varrho_{ij} = \varrho \,(\mathcal{T}^{*}_{ij},\mathcal{T}^{DNS}_{ij})= \varrho \,(\mathcal{T}^{R}_{ij},\mathcal{T}^{DNS}_{ij})$, where $\mathcal{T}^{DNS}_{ij}$ denotes the SGS stress tensor obtained from the DNS data.

\begin{figure}[t]
	\centering
	\includegraphics[width=3.5in]{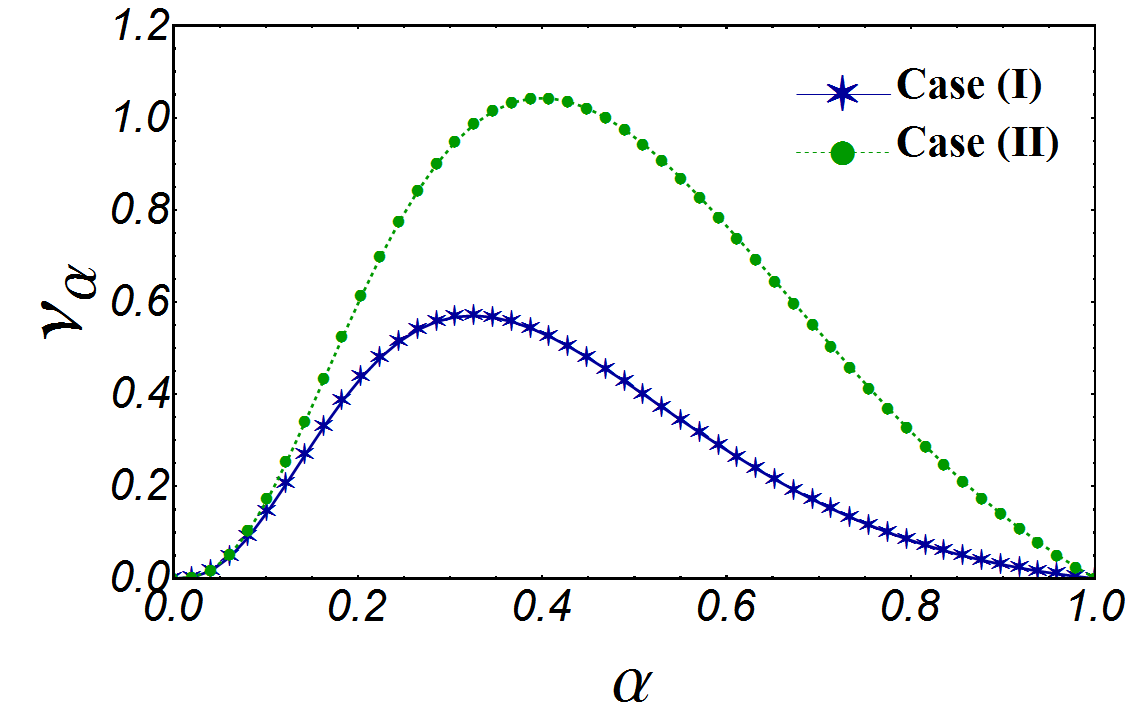}
	\caption{
		\label{fractional coeff}
		$\nu_{\alpha}$ versus $\alpha$ for $\nu=1.85 \times 10^{-4}$ in \textbf{Case (I)} and $\nu=10^{-3}$ in \textbf{Case (II)}}
\end{figure}

Technically, the proper choice of $\alpha$ can be made by looking at a range of $\alpha$, in which we obtain the relatively largest values of $\varrho_i$ while the corresponding $\mathcal{R}_i$ is around $1$. As a rule of thumb, $\mathfrak{C}_{\alpha}$ should be designed such that $\mathcal{R}_i\approx 1$ occurs, where the values of $\varrho_i$ are relatively maximum. With this in mind, we adopt $\mathfrak{C}_{\alpha}= \bar{c} \alpha^2$, where $\bar{c}=1500$. Figure \ref{fractional coeff} illustrates the variation of $\nu_{\alpha}$ versus $\alpha \in [0,1]$ for the specified properties of \textbf{Case (I)} and \textbf{Case (II)} in Tables \ref{Table: properties} and \ref{Table: C2-properties} at room temperature.

\begin{figure}[t!]
	\centering
	\begin{subfigure}[b]{0.32\textwidth}
		\centering
		\includegraphics[width=2.1in]{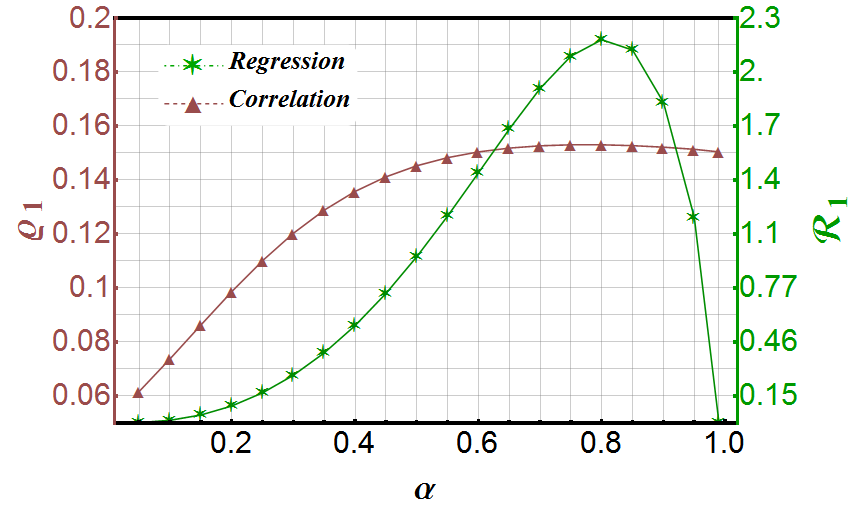}
		\subcaption*{}
	\end{subfigure}
	\begin{subfigure}[b]{0.32\textwidth}
		\centering
		\includegraphics[width=2.1in]{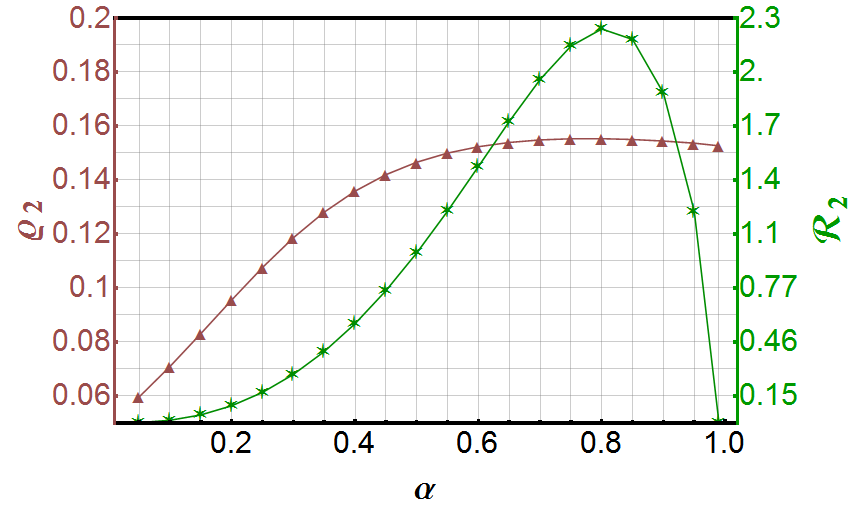}
		\subcaption{$\mathcal{L}_{\delta} = 4$}
	\end{subfigure}
	\begin{subfigure}[b]{0.32\textwidth}
		\centering
		\includegraphics[width=2.1in]{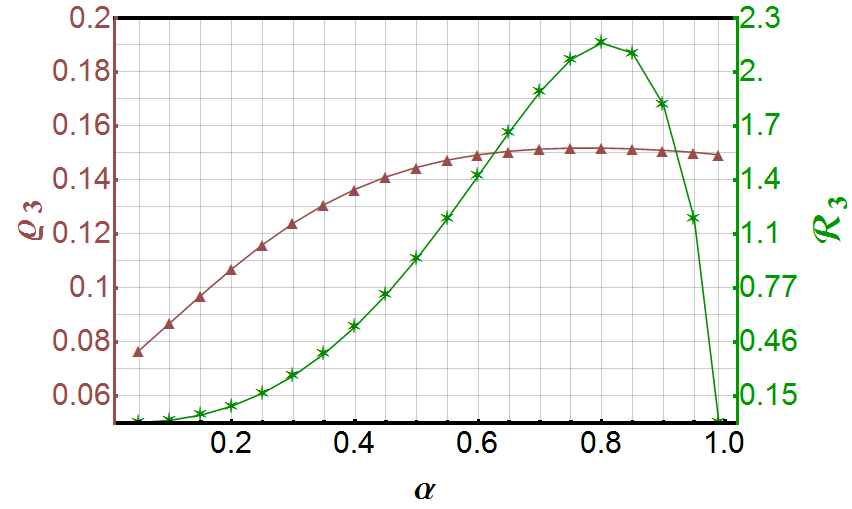}
		\subcaption*{}
	\end{subfigure}
	
	\vspace{0.2in}
	\begin{subfigure}[b]{0.32\textwidth}
		\centering
		\includegraphics[width=2.1in]{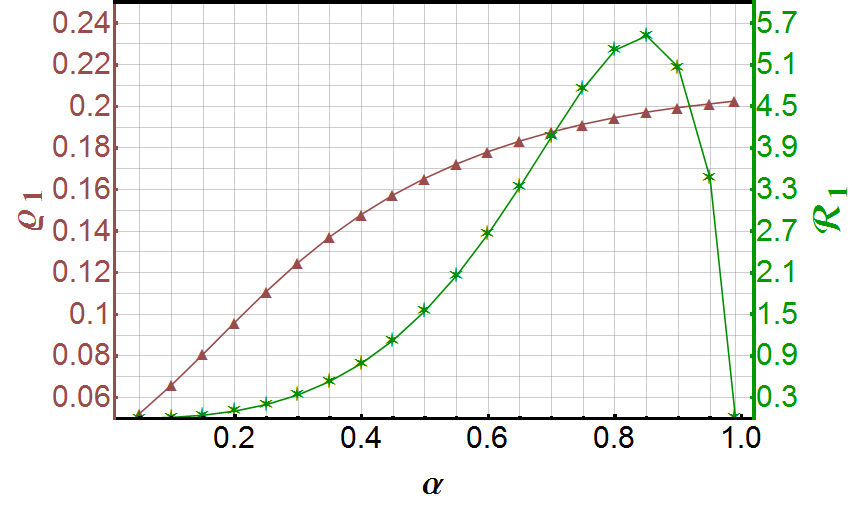}
		\caption*{$\mathcal{L}_{\delta} = 2 $}
		\subcaption*{}
	\end{subfigure}
	\begin{subfigure}[b]{0.32\textwidth}
		\centering
		\includegraphics[width=2.1in]{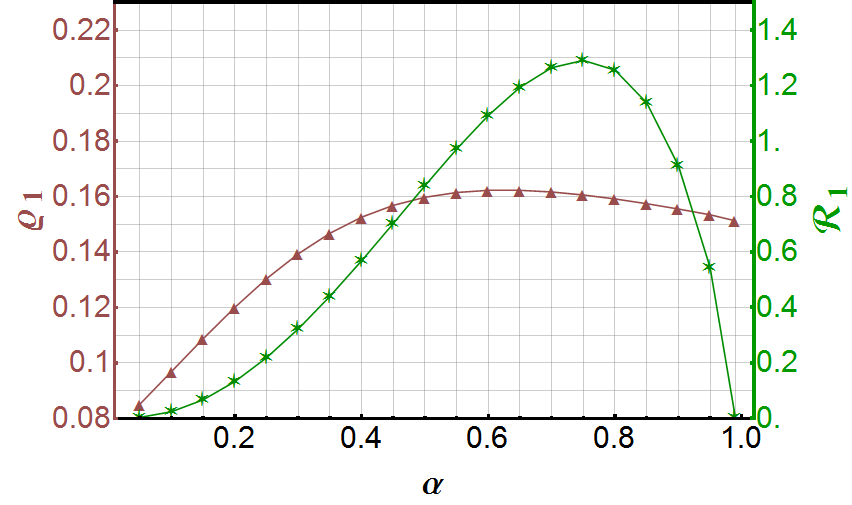}
		\caption*{$\mathcal{L}_{\delta} = 8$}
		\subcaption{}
	\end{subfigure}
	\begin{subfigure}[b]{0.32\textwidth}
		\centering
		\includegraphics[width=2.1in]{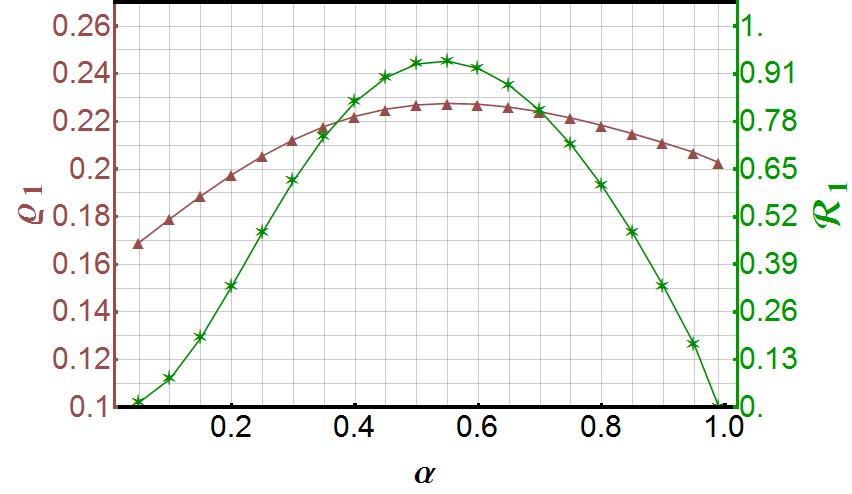}
		\caption*{$\mathcal{L}_{\delta}  = 32$}
		\subcaption*{}
	\end{subfigure}
	\caption{
		\label{fig: reg-corr-1}
		Variation of the correlation coefficient, $\varrho_i$, denoted by \textcolor{Mahogany}{$\blacktriangle$}, and the regression coefficient, $\beta_i$, denoted by \textcolor{Green}{\ding{86}}, in terms of the fractional exponent, $\alpha \in (0,1)$ using the DNS database of  \textbf{Case (I)} for (a) $i=1,2,3$ at $\mathcal{L}_{\delta} = 4$, and (b) $i=1$ at $\mathcal{L}_{\delta}  = 2,8,32$}
\end{figure}

\begin{figure}[t]
	\centering
	\includegraphics[width=2.9in]{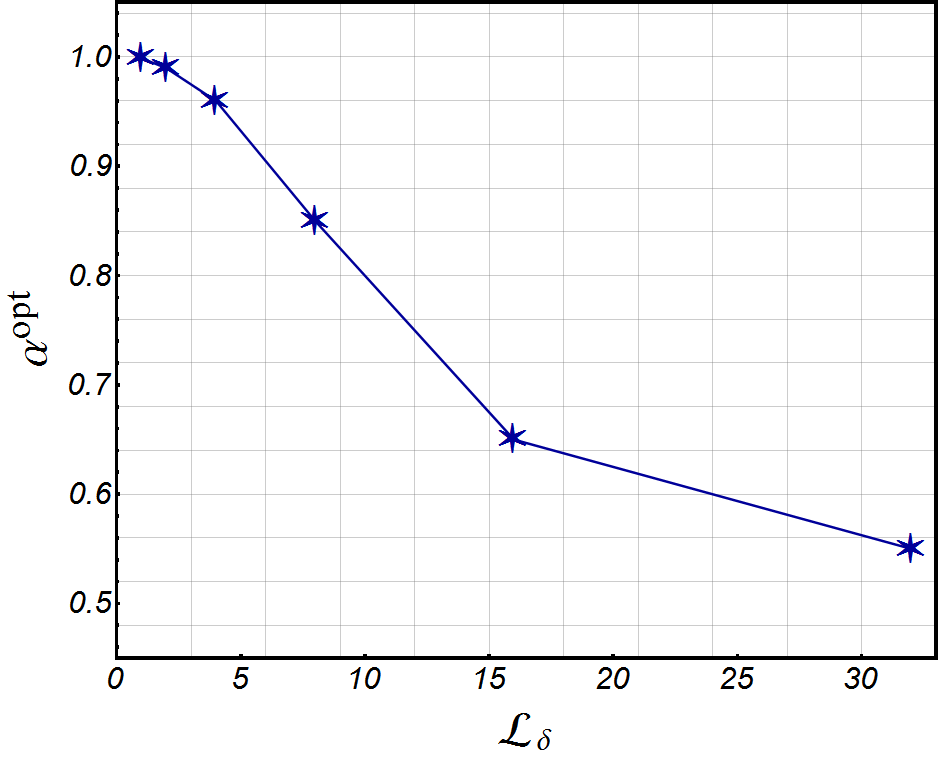}
	\caption{
		\label{optimum alpha}
		$\alpha^{opt}$ versus $\mathcal{L}_\delta$ for the \textbf{Case (I)}, of properties are given in Table \ref{Table: properties}}
\end{figure}

To estimate the optimal fractional exponent, $\alpha^{opt}$, in the case of forced HIT problem (\textbf{Case (I)}), we perform a comparative study of $\varrho_i$ and $\mathcal{R}_i$ versus $\mathcal{L}_{\delta}$ by carrying out several \textit{a priori} tests. 
In Figure \ref{fig: reg-corr-1}a, we illustrate $\varrho_i$ and $R_i$ for uniformly distributed $\alpha \in [0,1]$ at the specific $\mathcal{L}_{\delta} = 8$ for $i=1,2,3$. It is important to note that the values of $\alpha^{opt}$, obtained from the evaluations in each direction, can be approximately represented by the same value. Accordingly, we reduce the evaluation of $\alpha^{opt}$ to only the first direction as presented in Figure \ref{fig: reg-corr-1}b. After running enough test cases, we show the variations of $\alpha^{opt}$ versus $\mathcal{L}_{\delta}$ in Figure \ref{optimum alpha}. It reveals that enlarging $\mathcal{L}_{\delta}$ accelerates the reduction of $\alpha^{opt}$ toward the smaller values. Recall that $Re_{\lambda}$ remains approximately unchanged over time in forced HIT problems, hence, $\alpha^{opt}$ is primarily relying on $\mathcal{L}_{\delta}$. 

\begin{figure}[t!]
	\centering
	\begin{subfigure}[b]{0.45\textwidth}
		\centering
		\includegraphics[width=1 \linewidth ]{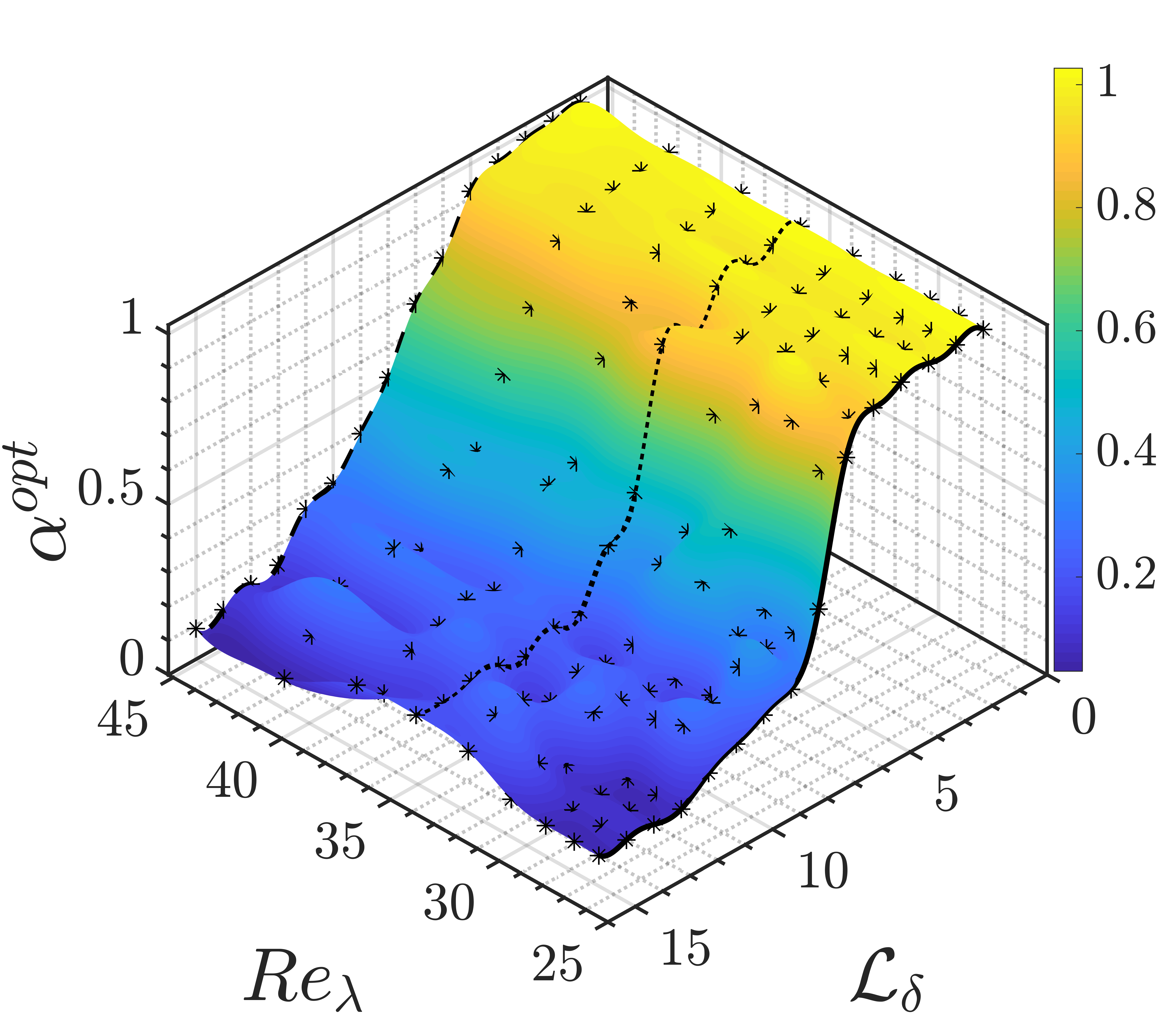}	
		\caption{}
		\label{optimum alpha-2a}
	\end{subfigure}
	\begin{subfigure}[b]{0.45\textwidth}
		\centering
		\includegraphics[width=1 \linewidth]{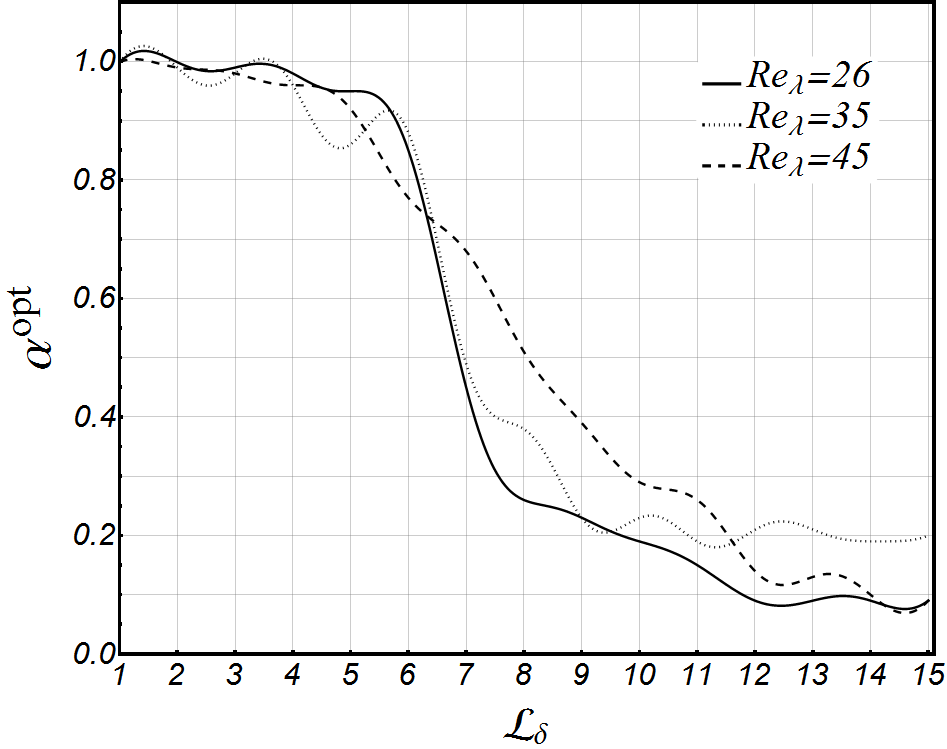}
		\caption{}
		\label{optimum alpha-2b}
	\end{subfigure}
	\caption{
		\label{optimum alpha-2}
		(a) The surface of $\alpha^{opt}$, obtained by the Kriging method, versus $\mathcal{L}_\delta$ and $Re_{\lambda}$ using the data points, denoted by $\star$, which are estimated by the \textit{a priori} tests of FSGS model for \textbf{Case (II)} and (b) comparison between the curves of $\alpha^{opt}$ versus $\mathcal{L}_\delta$, which are designated by $Re_{\lambda} \approx 26,35,45 $ }
\end{figure}

In a similar fashion, we perform \textit{a priori} tests of the FSGS model using the dataset of \textbf{Case (II)} for the purpose of calibrating $\alpha^{opt}$ to well-describe the non-Gaussian features of the SGS stresses. Such analysis also provides a platform for studying the statistical behavior of the FSGS model regarding a range of $Re_{\lambda}$. Using the Kriging method \cite{GEK_2014} from 135 direct evaluations, we approximate a high-resolution surrogate of $\alpha^{opt}$, which is presented as a function of $\mathcal{L}_{\delta}$ and $Re_{\lambda}$ in Figure \ref{optimum alpha-2a}. For three specific $Re_{\lambda}$, we also show the curves of $\alpha^{opt}$ versus $\mathcal{L}_{\delta}$ in Figure \ref{optimum alpha-2b}. Similar to the corresponding Figure in \textbf{Case (I)}, $\alpha^{opt}$ shows a substantial reduction by enlarging $\mathcal{L}_{\delta}$. Additionally, Figure \ref{optimum alpha-2b} confirms that, when $Re_{\lambda}$ decreases in a decaying process, $\alpha^{opt}$ exhibits a sharp reduction in a more limited span of $\mathcal{L}_{\delta}$. In further discussions, we elaborate on the results and the nonlocality effects induced by larger $Re_{\lambda}$ on $\alpha^{opt}$. 

\subsection{\textbf{Analysis of the Model Performance}}
\begin{figure}[t!]
	\centering
	\begin{subfigure}[b]{0.32\textwidth}
		\centering
		\includegraphics[height=2.2 in, width=2.05in]{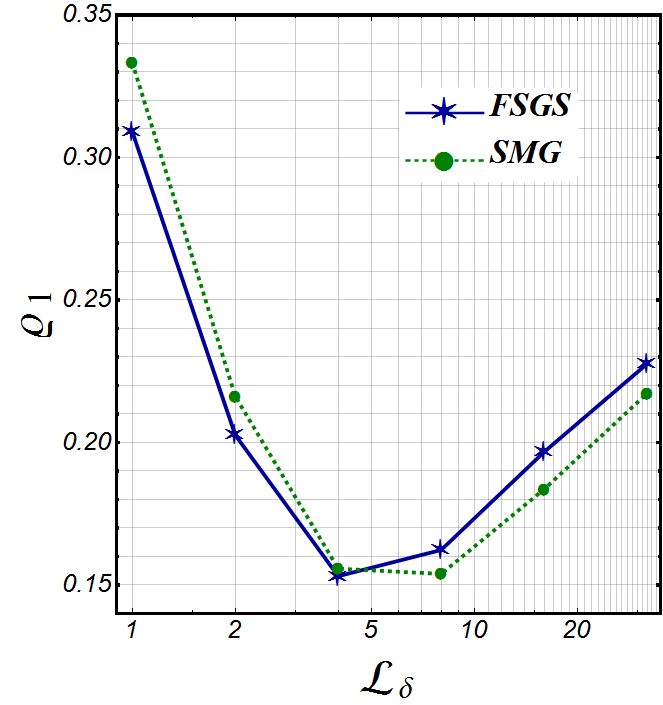}
		\caption*{}
	\end{subfigure}
	\begin{subfigure}[b]{0.32\textwidth}
		\centering
		\includegraphics[height=2.2 in, width=2.05in]{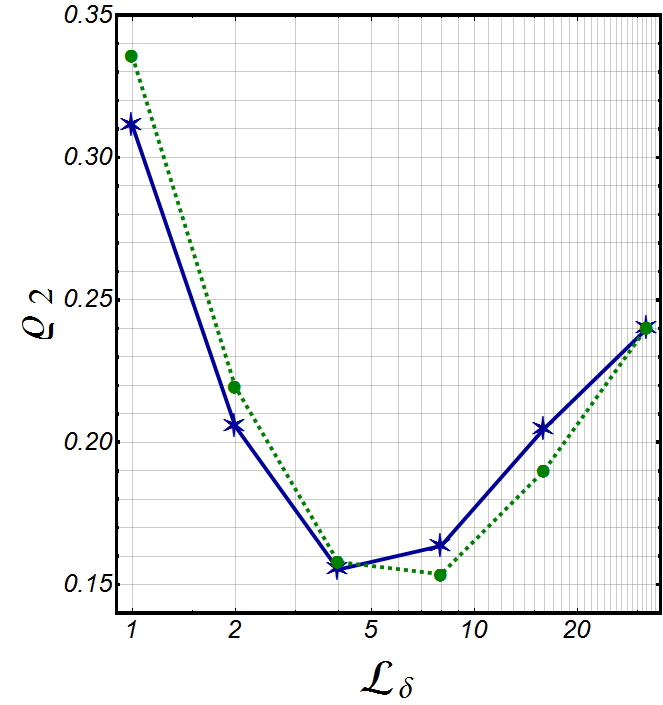}
		\caption*{}
	\end{subfigure}
	\begin{subfigure}[b]{0.32\textwidth}
		\centering
		\includegraphics[height=2.2 in, width=2.05in]{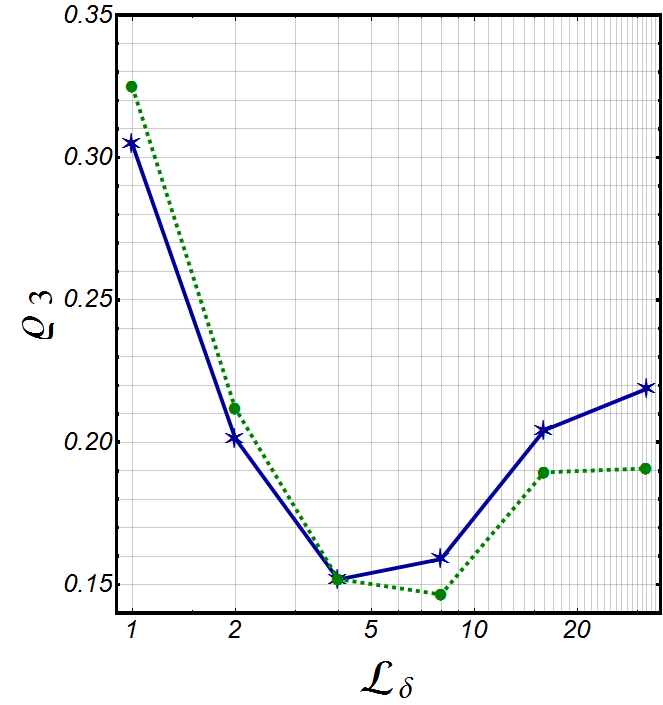}
		\caption*{}
	\end{subfigure}
	\setlength{\abovecaptionskip}{-10pt}
	\caption{
		\label{SMG-FSGS}
		Comparing the correlation coefficients, $\varrho_{i}$, from the FSGS model using the optimum fractional exponent, which is denoted by \textcolor{Blue}{\ding{86}}, with the corresponding ones implied by the Smagorinsky model, specified by \textcolor{Green}{\CIRCLE} }
\end{figure}

\begin{figure}[t!]
	\begin{subfigure}[b]{0.43\textwidth}
		\centering
		
		\includegraphics[width=2.56in]{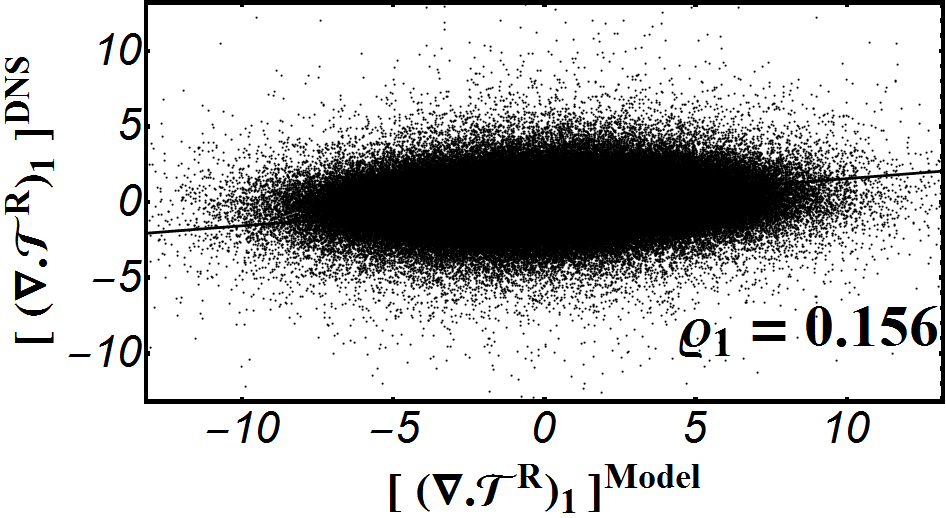}
		
		\subcaption*{$\mathcal{L}_{\delta} = 4$}
		
		\includegraphics[width=2.52in]{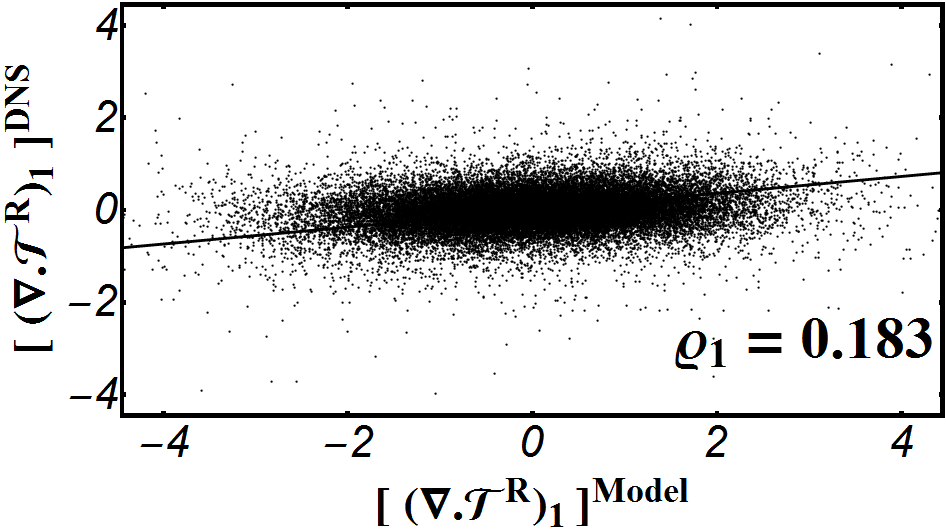}
		%
		
		\subcaption*{$\mathcal{L}_{\delta} = 16$}
	\end{subfigure}
	\begin{subfigure}[b]{0.43\textwidth}
		\centering
		
		\includegraphics[width=2.6in]{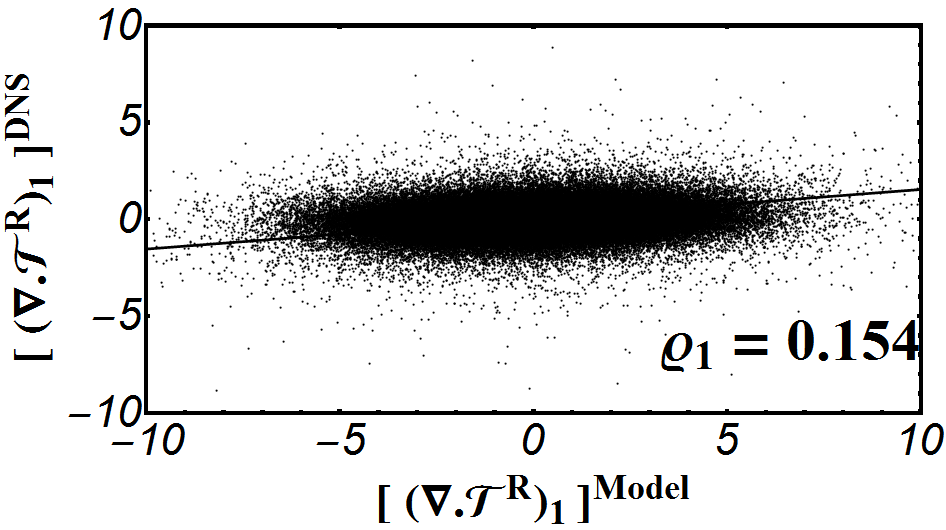}
		
		\subcaption*{$\mathcal{L}_{\delta} = 8$}
		
		\includegraphics[width=2.52in]{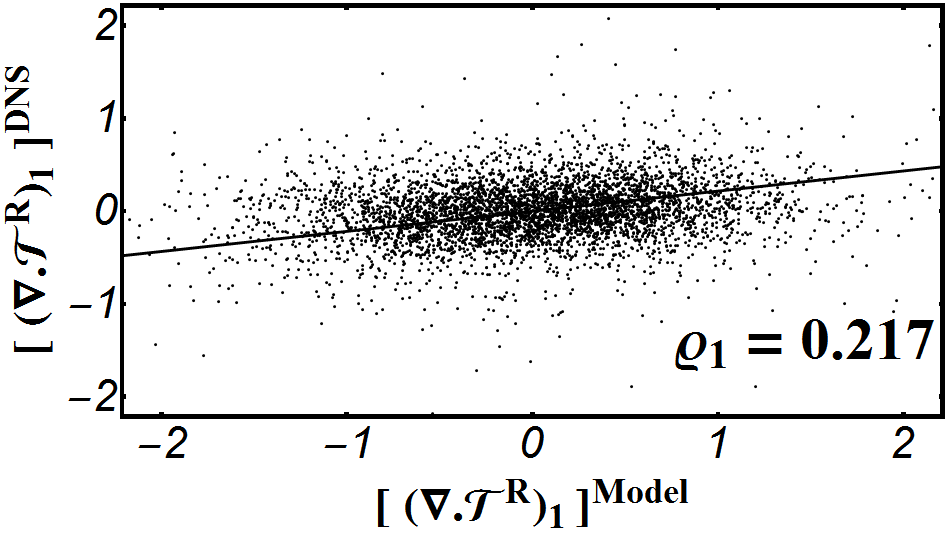}
		%
		
		\subcaption*{$\mathcal{L}_{\delta} = 32$}
	\end{subfigure}

	\caption{ 
		\label{scatter-plot}	
		\textit{A priori} results for the correlation between the true and model values for the components of $\nabla \cdot \mathcal{T}^{R}$, where $[\nabla \cdot \mathcal{T}^{R}]^{FSGS}= \mu_{\alpha} \, (-\Delta)^{\alpha} \bar{\textbf{V}} \vert_{\alpha=\alpha^{opt}} $, yielding the correlation coefficients, as shown}
\end{figure}

Following the evaluation of $\alpha^{opt}$, we perform a comparative study of performance for the FSGS model employing the introduced correlation coefficients, i.e., $\varrho_{i}$ and $\varrho_{ij}$. Using the high resolution turbulent fields of \textbf{Case (I)}, we compare the variation of $\varrho_{i}$ obtained from the FSGS and SMG models in terms of $\mathcal{L}_{\delta}$ in Figure \ref{SMG-FSGS}. Seemingly, the FSGS model shows acceptable correlations with the true values obtained from the DNS data, which is the notion of adequate magnitude and phase agreement. More significantly, by intensifying nonlocality in the filtered velocity field through increasing $\mathcal{L}_{\delta}$, the FSGS model works relatively better in terms of capturing heavy-tailed behavior of the SGS stresses in all directions. In Figure \ref{scatter-plot}, we present the scatter plot analysis on the values of $(\nabla \cdot \mathcal{T}^{R})_i$ for $i=1,2,3$, attained by the FSGS model and the filtered DNS data, for $\mathcal{L}_{\delta} = 8, \, 64$. The results confirm that, with a proper selection of $\alpha^{opt}$ in the FSGS model, we can achieve an approximate unit regression, which represents the same level of magnitudes in the scatter plots. In order to study the influence of $Re_{\lambda}$ on the performance, we reiterate the evaluation of correlation coefficients at other instantaneous realization of DNS database for \textbf{Case (I)} by imposing the same $\alpha^{opt}$.
The outcomes in Table \ref{my-label2} are seemingly in an acceptable agreement with the results shown in Figure \ref{SMG-FSGS}. Therefore, in the LES of forced HIT problems, $\alpha^{opt}$ can be dealt with as a constant parameter since $\mathcal{L}_{\delta}$ is considered as a constant value.


Despite the limitations of fractional approaches in approximating the SGS stresses, our findings explicitly formulate $\varrho_{ij}$  by evaluating $ \mathcal{T}^{*}_{ij}$ in \eqref{stss-2} on the filtered velocity field. Consistent with the results discussed previously, Table \ref{Table: SGS stresses} reports the correlations of the components of SGS stress tensor, $\varrho_{ij}$, for $\mathcal{L}_{\delta}=16$ and $32$. More clearly, the results support compatible behavior of the FSGS model with the SMG model in the description of SGS stresses.

\begin{figure}[t!]
	\centering
	\includegraphics[width = 1.05 \linewidth]{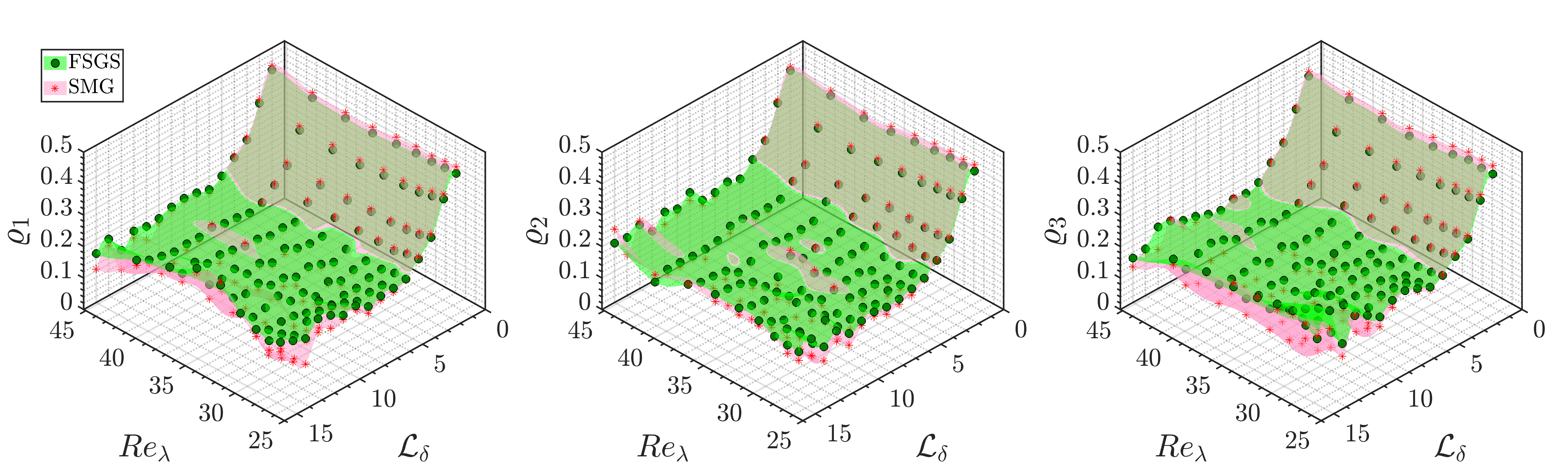}
	\setlength{\abovecaptionskip}{-1pt}
	\caption{
		\label{SMG-FSGS-2}
		Comparison of Kriging-constructed surfaces of $\varrho_{i}$ for $i=1,2,3$, obtained from \textit{a priori} study of the FSGS (denoted by \tikzcircle[fill=Green]{2.5pt}) and the SMG models (denoted by \textcolor{red}{$\ast$}), versus $\mathcal{L}_{\delta}$ and $Re_{\lambda}$ on \textbf{Case (II)}
	}
\end{figure}

In the LES of decaying HIT problems, $\alpha^{opt}$ retains $Re_{\lambda}$ dependence since $Re_{\lambda}$ as a macro-scale property undergoes a temporal decay. Employing $\alpha^{opt}$ in Figure \ref{optimum alpha-2}, we study the accuracy of the FSGS model in a broader framework, as shown in Figure \ref{SMG-FSGS-2}. Similar to \textbf{Case (I)},
the FSGS model shows better correlations at a wide range of $Re_{\lambda}$ by enlarging $\mathcal{L}_{\delta}$. Taken together, the FSGS model seems to be in a relatively favorable agreement with the filtered DNS database yet not comparable with structural models.


%

\begin{table}[t]
	\centering
	\caption{\textit{A priori} results for the correlation coefficients, $\varrho_{ij}$, of SGS stresses obtained from the FSGS and SMG models at two different Reynolds numbers, $\mathcal{L}_{\delta}$}
	\label{Table: SGS stresses}
	\vspace{-0.1 in}
	\begin{tabular}{c c c c c c c c c c c c c c c c}
		\hline \hline
		\multicolumn{1}{c}{  } 
		& \multirow{7}{*}{} &  \multicolumn{6}{c}{$\mathcal{L}_{\delta} = 16$} & &  \multicolumn{6}{c}{$\mathcal{L}_{\delta} = 32$}
		\\
		& &   $\varrho_{11}$ & $\varrho_{12}$ & $\varrho_{13}$& $\varrho_{22}$& $\varrho_{23}$& $\varrho_{33}$ & &   $\varrho_{11}$ & $\varrho_{12}$ & $\varrho_{13}$& $\varrho_{22}$& $\varrho_{23}$& $\varrho_{33}$
		\\
		\cline{1-1} \cline{3-8} \cline{10-15} FSGS & & $0.17$ &  $0.29$ & $0.30$ & $0.13$ & $0.30$ & $0.23$ & & $0.18$ &  $0.36$ & $0.33$ & $0.13$ & $0.33$ & $0.27$
		\\
		SMG & & $0.16$ &  $0.29$ & $0.28$ & $0.12$ & $0.31$ & $0.23$ & & $0.17$ &  $0.33$ & $0.32$ & $0.11$ & $0.32$ & $0.27$
	\end{tabular}
	\vspace{0.1 in}
\end{table}

\begin{table}[t!]
	\centering
	\caption{Study of FSGS model in terms of $\mathcal{L}$ through \textit{a priori} analysis at other time instants of \textbf{Case (I)}}
	\label{my-label2}
	\vspace{-0.1 in}	
	\begin{tabular}{c c c c c c c c c c c}
		\hline \hline
		\multicolumn{1}{c}{  } 
		& \multirow{5}{*}{} & \multicolumn{2}{c}{$Re_{\lambda} = 427$} & \multirow{5}{*}{} & \multicolumn{2}{c}{$Re_{\lambda} = 437$} & \multirow{5}{*}{} & \multicolumn{2}{c}{$Re_{\lambda} = 421$} 
		\\  
		\multicolumn{1}{c}{ $\mathcal{L}_{\delta}$ } 
		& & $4 $ &  $ 16$ &  & $ 4$ &  $ 16$ &   & $4 $ &  $ 16$  
		\\
		\cline{1-1}    \cline{3-4} \cline{6-7} \cline{9-10} $\varrho_{1}$  & & $0.15  $& $0.20  $ & & $ 0.15 $& $ 0.20$ & & $ 0.15 $& $ 0.20 $
		\\   $\varrho_{2}$ & & $0.16  $& $0.21 $ & & $ 0.15 $& $0.21$ & & $ 0.15$& $ 0.22 $
		\\   $\varrho_{3}$ & & $ 0.16  $& $0.20$ & & $0.15  $& $ 0.20$ & & $ 0.15 $& $0.21$
		\\ 
	\end{tabular}
\end{table}

\subsection{\textbf{Towards Modeling Nonlocal Effects}}

As pointed out previously, performance of the proposed model relies strictly on the selection of $\alpha^{opt}$ as a function of $Re_{\lambda}$ and $\mathcal{L}_{\delta}$. Regarding the connection between small scale turbulent motions in the NS and BT equations in section \ref{sec FBTE}, we explore the influence of nonlocal interactions on the model's performance at the microscopic level. Within the Boltzmann transport framework,  $\overline{f^{eq}(\Delta)}$ demonstrates increasingly multi-exponential behavior by enlarging $\mathcal{L}$. Practically, when we increase $\mathcal{L}$ in the filtered NS equations, more nonlocalities are incorporated into $\omega_{ij}$ in \eqref{GE-20} through $\overline{f^{eq}(\Delta)}$. From a physical point of view, vortices in turbulent flows tend to live longer than their turnover time. During the formation of coherent structures (see e.g., \cite{zayernouri2011coherent} and the references therein), the mutual advection and filamentation of vortices render nonlocal flow structures in isotropic turbulent flows. Filtering the flow field variables integrates such nonlocalities in a single numerical grid point, which intensifies the heavy-tailed characteristics of $\overline{f^{eq}(\Delta)}$. 

\begin{figure}[t!]
	\centering
	\begin{subfigure}[b]{ 0.4\textwidth}
		\centering
		\includegraphics[width=2.25in]{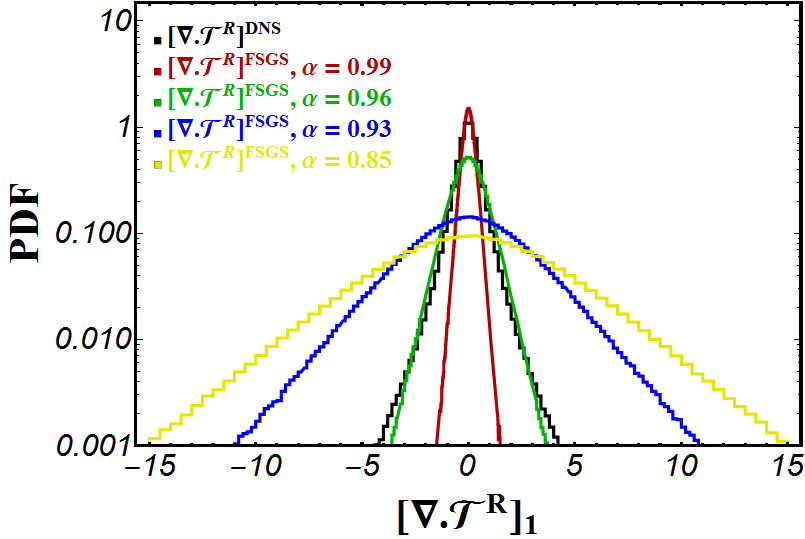}
		\caption*{$\mathcal{L}_{\delta} = 4$}
	\end{subfigure}
	%
	%
	\begin{subfigure}[b]{ 0.4\textwidth}
		\centering
		\includegraphics[width=2.25in]{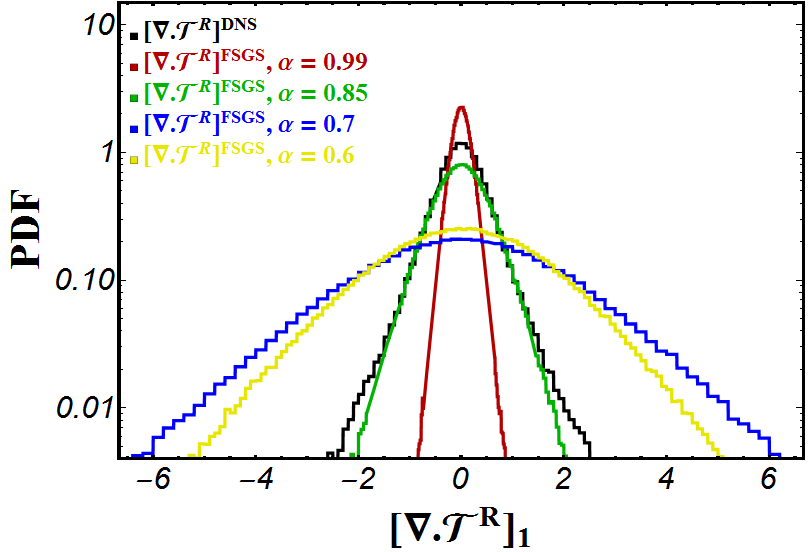}
		\caption*{$\mathcal{L}_{\delta} = 8$}
	\end{subfigure}
	\begin{subfigure}[b]{ 0.4\textwidth}
		\centering
		\includegraphics[width=2.25in]{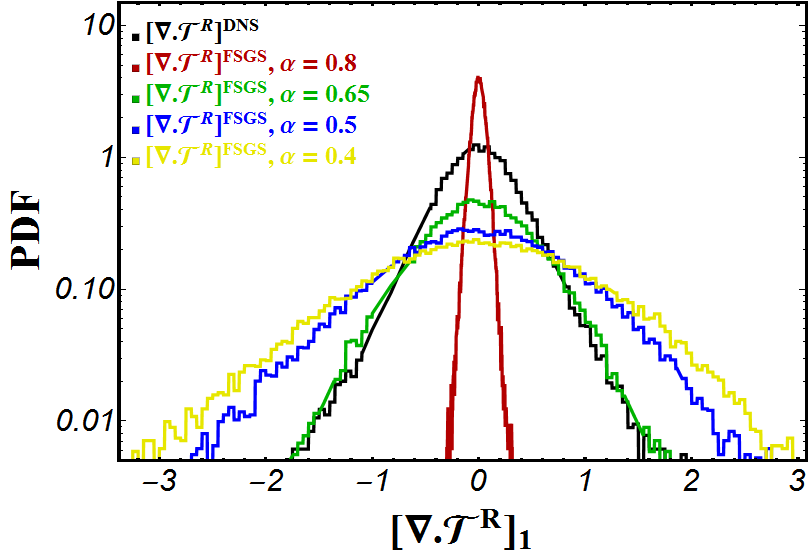}
		\caption*{$\mathcal{L}_{\delta} = 16$}
	\end{subfigure}
	\begin{subfigure}[b]{ 0.4\textwidth}
		\centering
		\includegraphics[width=2.25in]{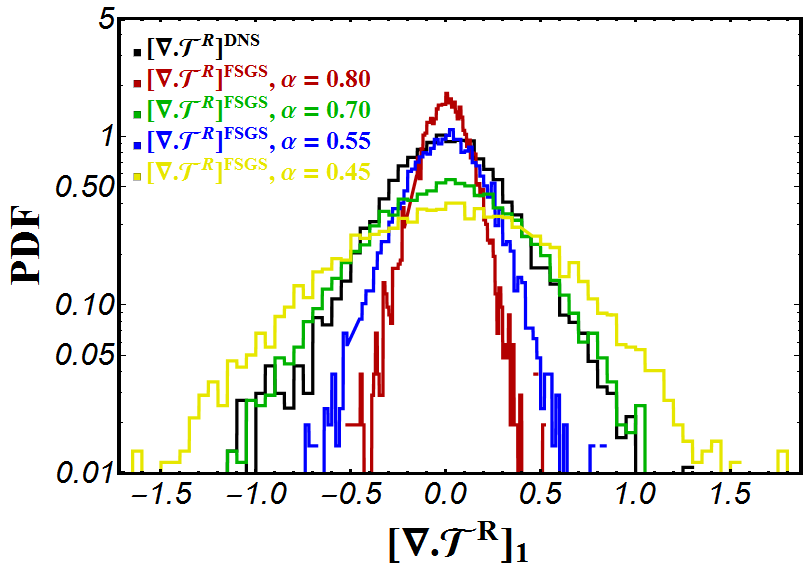}
		\caption*{$\mathcal{L}_{\delta} = 32$}
	\end{subfigure}
	
	\caption{ 
		\label{fig2: PDF-corr}	
		\textit{A priori} results for the PDF of the true and modeled $(\nabla \cdot \mathcal{T}^{R})_1$ regarding $\alpha$ variations at each $\mathcal{L}_{\delta}$}
\end{figure}

In the proposed framework, the multi-exponential behavior of $\overline{f^{eq}(\Delta)}$ is readily modeled by a \textit{L\'evy} $\beta$-stable distribution described in \eqref{GE-9-2}, in which the tail heaviness is indicated directly by $\mathcal{L}_{\delta}$. The multi-exponential pattern suggests that the heavy-tailed characteristics of $\overline{f^{eq}(\Delta)}$ get more intensified if we increase $\mathcal{L}_{\delta}$. Interestingly, as we decrease $\beta$, the \textit{L\'evy} $\beta$-stable distribution exhibits more fat-tailed behavior, which is provably in demand for the best-description of $\overline{f^{eq}(\Delta)}$. 
Extending this argument to the macroscopic level, the FSGS model inherently moves from the diffusion toward the advection to precisely represent the heavy-tailed behavior of SGS statistics by choosing the smaller values of $\alpha$ in \eqref{GE-7}. This argument accounts for the abrupt reduction of $\alpha^{opt}$ versus $\mathcal{L}_{\delta}$, presented in Figures \ref{optimum alpha} and \ref{optimum alpha-2}. For $\alpha<\alpha^{opt}$, the FSGS model is subject to overfit the heavy-tailed behavior of $\overline{f^{eq}(\Delta)}$, thereby losing the correlation. 

On such a background, the FSGS model can be dealt with as a new framework to capture anomalous features of SGS statistics at large values of $\mathcal{L}_{\delta}$. As shown in Figures \ref{SMG-FSGS} and \ref{SMG-FSGS-2}, the correlations associated with the FSGS model offer an improvement compared to the SMG model. Moreover, we proceed to perform qualitative assessment of the FSGS model in predicting the PDF of $(\nabla \cdot \mathcal{T}^{R})_i$ depicted in Figure \ref{fig2: PDF-corr} for different values of $\mathcal{L}_{\delta}$ in \textbf{Case (I)}. We should note that the presented results are confined to $i=1$ due to the similarities in other directions. It appears that by employing the proper choice of $\alpha^{opt}$, the PDF obtained by the FSGS model fits into the heavy-tailed distribution of true SGS values while with $\alpha < \alpha^{opt}$ the PDF is overpredicted. This argument emphasizes the reliability of the FSGS model on the selection $\alpha^{opt}$ as a function of $\mathcal{L}_{\delta}$ and $Re_{\lambda}$.  

Inevitably, due to the approximation we made in modeling $\overline{f^{eq}(\Delta)}$ in the filtered Boltzmann equation, there are some discrepancies between the results obtained from the fractional model and the filtered DNS data. We believe that this new framework has the advantage of allowing us to promote the accuracy of the model by involving more compatible options for approximating $\overline{f^{eq}(\Delta)}$ in \eqref{GE-7}. 




\section{Conclusion}
\label{Summary and Discussion}
%

This study presented a new framework to the functional modeling of SGS stresses in the LES of turbulent flows, starting from the kinetic theory. Within the proposed framework, we began with modeling the filtered equilibrium distribution function as a key term to consider the power-law scaling of SGS motions in the filtered BTE. Due to the multi-exponential behavior of the filtered equilibrium distribution function, we proposed to approximate it with a \textit{L\'evy}-stable distribution, where the associated fractional parameter strictly relied on the filter width. Subsequently, we derived the filtered NS equations from the approximated filtered BTE, in which the divergence of SGS stresses was modeled via a fractional Laplacian operator, $(-\Delta)^{\alpha}(\cdot) $ for $\alpha \in (0,1]$. In general, we established a framework, which permitted us to treat the source of turbulent motions at the kinetic level by employing a compatible choice of distribution function and derive the corresponding fractional operator in the filtered NS equations as an SGS model. Therefore, the proposed framework, termed ``FSGS modeling'', could potentially recover the non-Gaussian statistics of SGS motions precisely. Next, we studied the physical and mathematical properties of the proposed model and introduced a set of mild conditions to preserve the second law of thermodynamics. Eventually, we carried out \textit{a priori} evaluations of the FSGS model based on the DNS database of forced and decaying HIT problems. In light of the analysis, there was a relatively great agreement between the modeled and true SGS values in terms of the correlation and regression coefficients. The performance of the FSGS model depended rigorously on the choice of fractional exponent, $\alpha$, as a function of $\mathcal{L}$ and $Re_{\lambda}$. We showed that, by enlarging $\mathcal{L}$, the heavy-tailed characteristics of the SGS motions could become more intensified, which were conceivably well-described by the FSGS model with smaller values of $\alpha$. With all this in mind, FSGS modeling provided a new perspective, which respected the non-Gaussian behavior of SGS stresses by exploiting fractional calculus within the Boltzmann transport framework.



On the basis of the theoretical background and the \textit{a priori} analyses provided in this study, the proposed framework has a remarkable potential to outline more sophisticated SGS models in LES of turbulent flows by leveraging proper mathematical tools in fractional calculus. As a part of future works, the FSGS model can be enhanced in order
to achieve comparatively greater correlations regarding the structural SGS models. In further studies, we perform parameter calibration to determine the best performance of FSGS models and find the optimum combination of the underlying parameters. We also carry out \textit{a posteriori} analysis of FSGS models in an LES solver as an ultimate examination.

\section*{Appendix}
In this section, we follow the derivations of fractional NS equations in \cite{epps2018turbulence} to evaluate the shear and SGS stresses in \eqref{GE-26-2} and \eqref{GE-27}.
\subsection*{$\bullet$ \textbf{Temporal Shift}}
\label{Apx: temporal shift}
Recalling from the Assumption \ref{Rem-1} that $s \sim \mathcal{O}(1)$, we take the temporal Taylor expansion of $f^{*}$ as follows:
\begin{eqnarray}
\label{AP_1}
f_{s,s}^{*} = f^{*}\big (\bar{\Delta}(t-s\tau,\boldsymbol{x}-s\tau \boldsymbol{u}, \boldsymbol{u}) \big ) &=& f^{*}_s\big (\bar{\Delta} \big ) + \frac{\partial f^{*}_s}{\partial \bar{\Delta}} \frac{\partial \bar{\Delta}}{\partial t} \delta t + \mathcal{O} (\delta t^2)
\nonumber
\\
&=& f^{*}_s\big (\bar{\Delta} \big ) + \frac{\partial f^{*}_s}{\partial \bar{\Delta}} \frac{\partial \bar{\Delta}}{\partial t} (-s\tau) + \mathcal{O} (\tau^2),
\end{eqnarray}
where $f^{*}_s\big (\bar{\Delta} \big )=f^{*}\big (\bar{\Delta}(t,\boldsymbol{x}-s\tau \boldsymbol{u}, \boldsymbol{u}) \big )$, $\delta t = -s \tau$ and 
\begin{equation}
\label{AP_2}
\frac{\partial \bar{\Delta}}{\partial t} = \frac{-2}{U^2} \sum_{k=1}^{3} (u_k-\bar{V}_k) \frac{\partial \bar{V}_k}{\partial t}.
\end{equation} 
Considering \eqref{AP_1}, we can approximate \eqref{GE-19-1} according to
\begin{eqnarray}
\label{Ap_3}
\varsigma_{ij} &\approx& \int_{0}^{\infty} e^{-s} \int_{\mathbb{R}^d}(u_i-\bar{V}_i)(u_j-\bar{V}_j) \big ( f^{*}_s\big (\bar{\Delta} \big ) + \frac{\partial f^{*}_s}{\partial \bar{\Delta}} \frac{\partial \bar{\Delta}}{\partial t} (s\tau) \big ) d\boldsymbol{u} \, ds,
\\
\nonumber
&=& 
\int_{0}^{\infty} e^{-s} \int_{\mathbb{R}^d}(u_i-\bar{V}_i)(u_j-\bar{V}_j) \bigg [ f^{*}_s\big (\bar{\Delta} \big ) + \frac{2}{U^2} \frac{\partial f^{*}_s}{\partial \bar{\Delta}}\, (\sum_{k=1}^{3} (u_k-\bar{V}_k) \frac{\partial \bar{V}_k}{\partial t})\, (s\tau) \bigg ] d\boldsymbol{u} \, ds.
\end{eqnarray} 
Since $\bar{\Delta}$ is an even function of $(u_k-\bar{V}_k)$ for $k=1,\cdots,3$, $f^{eq}(\bar{\Delta})$ and $f^{Model}(\bar{\Delta})$ and also their corresponding first derivatives $\frac{\partial f^{eq}_s}{\partial \bar{\Delta}}$ and $\frac{\partial f^{Model}_s}{\partial \bar{\Delta}}$ are even functions of $(u_k-\bar{V}_k)$. Subsequently, there is an odd power of either $(u_i-\bar{V}_i)$ or $(u_j-\bar{V}_j)$, which makes $$\int_{\mathbb{R}^d}(u_i-\bar{V}_i)(u_j-\bar{V}_j) \bigg [  \frac{\partial f^{*}_s}{\partial \bar{\Delta}}\, (\sum_{k=1}^{3} (u_k-\bar{V}_k) \frac{\partial \bar{V}_k}{\partial t})\, (s\tau) \bigg ] d\boldsymbol{u} = 0.$$
Therefore,
\begin{eqnarray}
\label{Ap_4}
\varsigma_{ij} &\approx& \int_{0}^{\infty} e^{-s} \int_{\mathbb{R}^d}(u_i-\bar{V}_i)(u_j-\bar{V}_j) f^{*}_s\big (\bar{\Delta} \big )  d\boldsymbol{u} \, ds
\nonumber
\\
&=& \int_{\mathbb{R}^d} \int_{0}^{\infty} e^{-s} (u_i-\bar{V}_i)(u_j-\bar{V}_j) f^{*}_s\big (\bar{\Delta} \big )  ds \,  d\boldsymbol{u}.
\end{eqnarray} 
%

\subsection*{$\bullet$ \textbf{Shear Stresses}}
Regarding \eqref{GE-25}, the shear stress tensors are described according to
\begin{eqnarray}
\nonumber
\mathcal{T}_{ij}^{Shear} &=& \int_{0}^{\infty}  \int_{\mathbb{R}^d} (u_i-\bar{V}_i)(u_j-\bar{V}_j) (f^{eq}_{s}(\bar{\Delta})-f^{eq}(\bar{\Delta})) e^{-s} d\boldsymbol{u}\,  ds,
\end{eqnarray}
in which $f^{eq}_{s}(\bar{\Delta})=f^{eq}(\bar{\Delta}(t,\boldsymbol{x}-s\tau \boldsymbol{u}, \boldsymbol{u})$. The spatial shift $\delta x = s\tau \vert \boldsymbol{u} \vert$ can be decomposed into small $\delta x \leq l$ and large $\delta x > l$ displacements, which are associated with $ \bar{\Delta} \leq 1$ and $ \bar{\Delta} > 1$, respectively. Therefore,
\begin{eqnarray}
\label{AP_5-0}
\mathcal{T}_{ij}^{Shear} &=& \int_{0}^{\infty}  \int_{\delta x \leq l} (u_i-\bar{V}_i)(u_j-\bar{V}_j) (f^{eq}_{s}(\bar{\Delta})-f^{eq}(\bar{\Delta})) e^{-s} d\boldsymbol{u}\,  ds
\nonumber
\\
&&+ \int_{0}^{\infty}  \int_{\delta x > l} (u_i-\bar{V}_i)(u_j-\bar{V}_j) (f^{eq}_{s}(\bar{\Delta})-f^{eq}(\bar{\Delta})) e^{-s} d\boldsymbol{u}\,  ds.
\end{eqnarray}
Since $f^{eq}(\bar{\Delta})$ belongs to $C^{\infty}$, which denotes the space of infinitely differentiable functions, we can perform the local linear approximation of $f^{eq}_{s}(\bar{\Delta})$, which yields in 
\begin{equation}
\label{AP_5-1}
f^{eq}_{s}(\bar{\Delta})\approx f^{eq}(\bar{\Delta}) +  \frac{\partial f^{eq}}{\partial \bar{\Delta}} (\bar{\Delta}_s - \bar{\Delta}),
\end{equation}
where $ \frac{\partial f^{eq}}{\partial \bar{\Delta}} = -\frac{\rho}{2 U^3} e^{-\bar{\Delta}/2}$ and $\bar{\Delta}_s = \bar{\Delta}(t,\boldsymbol{x}-s\tau \boldsymbol{u}, \boldsymbol{u})$. Due to the exponential behavior of $f^{eq}_{s}(\bar{\Delta})-f^{eq}(\bar{\Delta})$, we obtain
\begin{eqnarray}
\label{AP_5-2}
\int_{0}^{\infty}  \int_{\delta x > l} (u_i-\bar{V}_i)(u_j-\bar{V}_j) (f^{eq}_{s}(\bar{\Delta})-f^{eq}(\bar{\Delta})) e^{-s} d\boldsymbol{u}\,  ds \approx 0
\end{eqnarray}
and thereby
\begin{eqnarray}
\label{AP_5-3}
\mathcal{T}_{ij}^{Shear} &\approx& \int_{0}^{\infty}  \int_{\delta x \leq l} (u_i-\bar{V}_i)(u_j-\bar{V}_j) (f^{eq}_{s}(\bar{\Delta})-f^{eq}(\bar{\Delta})) e^{-s} d\boldsymbol{u}\,  ds.
\end{eqnarray}
Moreover, it is permissible to use the Taylor expansion of  $\bar{\Delta}_s$ for $\delta x \leq l$, which is formulated as $$\bar{\Delta}_s = \bar{\Delta} + \frac{\partial \bar{\Delta}}{\partial x_k} \delta x_k  + \mathcal{O} \big (\vert \delta x \vert^2 \big ),$$
where
\begin{eqnarray}
\label{AP_6}
\frac{\partial \bar{\Delta}}{\partial x_k} = \frac{-2}{U^2} \sum_{m=1}^{3} (u_m-\bar{V}_m) \frac{\partial \bar{V}_m}{\partial x_k}.
\end{eqnarray}
Therefore,
\begin{eqnarray}
\label{AP_5}
f_{s}^{eq} = f^{eq}\big (\bar{\Delta}(t,\boldsymbol{x}-s\tau \boldsymbol{u}, \boldsymbol{u}) \big ) &=& f^{eq}\big (\bar{\Delta} \big ) + \frac{\partial f^{eq}}{\partial \bar{\Delta}} \frac{\partial \bar{\Delta}}{\partial x_k} \delta x_k + \mathcal{O} \big (\vert\delta x \vert^2 \big )
\nonumber
\\
&=& f^{eq}\big (\bar{\Delta} \big ) + \frac{\partial f^{eq}}{\partial \bar{\Delta}} \frac{\partial \bar{\Delta}}{\partial x_k} (-s\tau u_k) + \mathcal{O} \big ((s\tau \vert\boldsymbol{u} \vert)^2 \big ).
\end{eqnarray}
Plugging \eqref{AP_5} and \eqref{AP_6} into \eqref{GE-25}, we attain 
\begin{eqnarray}
\label{AP_7}
\mathcal{T}_{ij}^{Shear} &\approx& - \int_{0}^{\infty}  \int_{\delta x \leq l} (u_i-\bar{V}_i)(u_j-\bar{V}_j) \bigg ( \frac{\partial f^{eq}}{\partial \bar{\Delta}} \frac{\partial \bar{\Delta}}{\partial x_k} (s\tau u_k) \bigg ) e^{-s} d\boldsymbol{u}\,  ds
\nonumber
\\
&=&
-\frac{2\rho}{U^5}\int_{0}^{\infty}  \int_{\delta x \leq l} (u_i-\bar{V}_i)(u_j-\bar{V}_j) \bigg ( e^{-\Delta}\, \sum_{m=1}^{3} (u_m-\bar{V}_m) \frac{\partial \bar{V}_m}{\partial x_k} \bigg ) (s\tau u_k)e^{-s} d\boldsymbol{u}\,  ds.
\nonumber
\end{eqnarray}
We should note that the limits of integral in \eqref{AP_7} can be extended to $\mathbb{R}^3$ due to \eqref{AP_5-2}. Following every steps in the derivation of shear stresses from (4.33) to (4.36) in \cite{epps2018turbulence}, we can formulate $\mathcal{T}_{ij}^{Shear}=\mu \Big ( \frac{\partial \bar{V}_i}{\partial x_j} + \frac{\partial \bar{V}_j}{\partial x_i}  \Big )$ in \eqref{GE-26-2} from \eqref{AP_7}, in which we obtain $\mu = \frac{-2 \rho \tau}{U^5} \int_{0}^{\infty} I_0 s e^{-s} ds = \rho U^2 \tau$ and $I_0 = \frac{4\pi}{15} \int_{0}^{\infty} r^6 e^{-\Delta} dr$, where $\Delta = \frac{r^2}{U^2}$ and $r = \vert \boldsymbol{u} - \boldsymbol{\bar{V}}\vert$. 

\subsection*{$\bullet$ \textbf{SGS Stresses}}
The SGS stresses are given by
\begin{eqnarray}
\label{AP_8}
\mathcal{T}_{ij}^{R} = \int_{0}^{\infty}  \int_{\mathbb{R}^d} (u_i-\bar{V}_i)(u_j-\bar{V}_j) (f^{\beta}_{s}(\bar{\Delta})-f^{\beta}(\bar{\Delta})) e^{-s} d\boldsymbol{u}\,  ds
\end{eqnarray}
in \eqref{GE-26}, where $f^{\beta}(\bar{\Delta}) = \frac{\rho}{U^3}F^{\beta}(\bar{\Delta})$ and $F^{\beta}(\bar{\Delta})$ denotes the isotropic \textit{L\'evy}-$\beta$ stable distribution. Therefore,
\begin{equation}
\label{AP_9}
\mathcal{T}_{ij}^{R} = -\frac{\rho}{U^3} \int_{0}^{\infty}  \int_{\mathbb{R}^d} (u_i-\bar{V}_i)(u_j-\bar{V}_j) (F^{\beta}_{s}(\bar{\Delta})-F^{\beta}(\bar{\Delta})) e^{-s} d\boldsymbol{u}\,  ds.
\end{equation}
Asymptotically, $F^{\beta}(\bar{\Delta})$ behaves like a power-law distribution when $\bar{\Delta} > 1$, i.e., $F^{\beta}(\bar{\Delta}) \sim \tilde{C}_{\beta} \bar{\Delta}^{\beta}= \frac{C_{\alpha}}{ \bar{\Delta}^{-\alpha+d/2}}$, where $\beta = -\alpha - \frac{d}{2}$ and $C_{\alpha}=\frac{2^{2\alpha} \Gamma(\alpha+d/2)}{\pi^{d/2} \Gamma(-\alpha)}$. It is worth mentioning that $\overline{f^{eq}(\Delta)}$ demonstrates a heavy-tailed behavior at $\bar{\Delta} > 1$, it keeps the exponential trait for $\bar{\Delta} < 1$ though. Regarding the exponential behavior of $f^{eq}(\bar{\Delta})$, $\overline{f^{eq}(\Delta)}-f^{eq}(\bar{\Delta})$ can be fitted by a heavy-tailed distribution like $F^{\beta}(\bar{\Delta})$, in which $F^{\beta}(\bar{\Delta})$ reduces exponentially in a close proximity of $\bar{\Delta} = 0$. Therefore, we can simplify \eqref{AP_9} to
\begin{eqnarray}
\label{AP_10}
\mathcal{T}_{ij}^{R} &\approx& -\frac{\rho C_{\alpha}}{U^3}  \int_{0}^{\infty}  \int_{\mathbb{R}^d-B_{\epsilon}} (u_i-\bar{V}_i)(u_j-\bar{V}_j) (\bar{\Delta}^{-\alpha+d/2}_s-\bar{\Delta}^{-\alpha+d/2}) e^{-s} d\boldsymbol{u}\,  ds,
\end{eqnarray}
where $d = 3$ and $B_{\epsilon}=\{ u\in \mathbb{R}^d\, s.t. \, \vert \bar{\Delta} \vert < \epsilon \}$, which is associated with $\bar{\Delta} \ll 1$. Due to the fact that $F^{\beta}(\bar{\Delta})$ is continuously differentiable for $\vert \boldsymbol{u} \vert \in \mathbb{R}^d-B_{\epsilon} $, we perform the Taylor expansion of $F^{\beta}_s(\bar{\Delta})$ as follows:
$$ F^{\beta}_s(\bar{\Delta})-F^{\beta}(\bar{\Delta}) \approx \frac{\partial F^{\beta}(\bar{\Delta})}{\partial \bar{\Delta}} (\bar{\Delta}_s-\bar{\Delta}) = (\alpha+\frac{3}{2}) \, \frac{\rho C_{\alpha} }{U^3}\, \frac{(\bar{\Delta}_s-\bar{\Delta})}{\bar{\Delta}^{\alpha+5/2}}. $$
Under Assumption \ref{Rem-1}, in which $s\sim \mathcal{O}(1)$, we obtain
$$ \delta x = s \vert \boldsymbol{u} \vert \tau > \mathcal{O}(l) \rightarrow \vert \boldsymbol{u} \vert > \mathcal{O}(\frac{l}{\tau}) = \mathcal{O}(\frac{l}{\lambda/U}) > \mathcal{O}(U) $$
at large $\delta x > l$, which yields in $\bar\Delta = \frac{\vert \boldsymbol{u} - \bar{\boldsymbol{V}} \vert^2 }{U^2} \approx \frac{\vert \boldsymbol{u} \vert^2 }{U^2} \gg 1$ and $u_i -\bar{V}_i \approx u_i$. In virtue of (4.50-51) in \cite{epps2018turbulence}, we also conclude that 
\begin{equation}
\label{AP_10_1}
\bar{\Delta}_s-\bar{\Delta} \approx -2 \sum_{k=1}^{3}\frac{ u_k(\bar{V}_k(\boldsymbol{x})-\bar{V}_k(\boldsymbol{x}^{\prime}))}{U^2}.
\end{equation}
Utilizing the definition of $\boldsymbol{u} = \frac{\boldsymbol{x} - \boldsymbol{x}^{\prime}}{s\, \tau}$ in Section \ref{sec 4.1} and \eqref{AP_10_1}, we reformulate \eqref{AP_10} as 
\begin{eqnarray}
\label{AP_11}
\mathcal{T}_{ij}^{R} &\approx& (\alpha+\frac{3}{2}) \frac{\rho C_{\alpha}}{U^3} \int_{0}^{\infty}  \int_{\mathbb{R}^d-B_{\epsilon}} (\frac{x_i -x_i^{\prime}}{s\tau}) \, (\frac{x_j -x_j^{\prime}}{s\tau}) \frac{(\bar{\Delta}_s-\bar{\Delta})}{(\frac{ \vert \boldsymbol{x} -\boldsymbol{x}^{\prime}\vert}{s\tau U})^{2\alpha+5}}     \frac{d\boldsymbol{x}^{\prime}}{(s\tau)^3}\,  e^{-s}  ds,
\nonumber
\\
\nonumber
&=& (2\alpha+3) (\rho C_{\alpha} \tau^{2\alpha-1} U^{2\alpha}) \int_{0}^{\infty}  \frac{e^{-s}}{s^{1-2\alpha}}  ds \times 
\\
&&\int_{\mathbb{R}^d-B_{\epsilon}}  (x_i -x_i^{\prime}) \, (x_j -x_j^{\prime}) \frac{(\boldsymbol{x} -\boldsymbol{x}^{\prime})\cdot (\bar{\boldsymbol{V}}(\boldsymbol{x})- \bar{\boldsymbol{V}} (\boldsymbol{x}^{\prime}) ) }{\vert \boldsymbol{x} -\boldsymbol{x}^{\prime}\vert^{2\alpha+5}}     d\boldsymbol{x}^{\prime},
\end{eqnarray}
which corresponds to (4.58) in \cite{epps2018turbulence}. Therefore, we can proceed the same derivations as discussed in (4.58) to (4.64) in \cite{epps2018turbulence} to obtain
\begin{eqnarray}
\label{AP_12}
(\nabla \cdot \mathcal{T}^{R})_i&=&\frac{\rho (U\tau)^{2\alpha}}{\tau}\Gamma(2\alpha+1) \, C_{\alpha} \int_{\mathbb{R}^d-B_{\epsilon}} \frac{\bar{V}_i(\boldsymbol{x}^{\prime})-\bar{V}_i(\boldsymbol{x})}{\vert \boldsymbol{x}^{\prime}-\boldsymbol{x} \vert^{2\alpha+d}}d\boldsymbol{x}^{\prime}
\nonumber
\\
&=& p.v. \frac{\rho (U\tau)^{2\alpha}}{\tau}\Gamma(2\alpha+1) \, C_{\alpha} \int_{\mathbb{R}^d} \frac{\bar{V}_i(\boldsymbol{x}^{\prime})-\bar{V}_i(\boldsymbol{x})}{\vert \boldsymbol{x}^{\prime}-\boldsymbol{x} \vert^{2\alpha+d}}d\boldsymbol{x}^{\prime}
\end{eqnarray}
in which "p.v." denotes the principal value of the integral.

\newpage
\bibliographystyle{siam}
\bibliography{RFSLP_Refs3}

\end{document}